\documentclass[times,twocolumn,final,nopreprintline]{elsarticle}
\usepackage{amsmath,amssymb,amsfonts}
\usepackage{graphicx}
\usepackage{booktabs,makecell,longtable}
\usepackage{latexsym}
\usepackage{multirow}
\usepackage{url}
\usepackage{float}
\usepackage{csquotes}
\usepackage{enumitem}
\usepackage[dvipsnames]{xcolor}
\usepackage{subcaption}
\usepackage{caption}

\usepackage{tabularx}
\usepackage{tabulary}

\usepackage{xtab,afterpage}
\definecolor{newcolor}{rgb}{.8,.349,.1}

\journal{Medical Image Analysis}

\begin{document}

\begin{frontmatter}

%\title{A comprehensive survey of deep learning methods in computational pathology: Application to prediction and prognosis}%
\title{Deep neural network models for computational histopathology: A survey}%

\author[a,b]{Chetan L. Srinidhi \corref{cor1}}
\author[b]{Ozan Ciga}
\author[a,b]{Anne L. Martel}
\cortext[cor1]{Corresponding author: \\
\textit{E-mail address:} chetan.srinidhi@utoronto.ca (Chetan L. Srinidhi) \\ \textit{Published in Medical Image Analysis, Vol. 67, Jan 2021, Elsevier (https://doi.org/10.1016/j.media.2020.101813)}}
 
\address[a]{Physical Sciences, Sunnybrook Research Institute, Toronto, Canada}
\address[b]{Department of Medical Biophysics, University of Toronto, Canada}

\begin{abstract}
Histopathological images contain rich phenotypic information that can be used to monitor underlying mechanisms contributing to disease progression and patient survival outcomes. Recently, deep learning has become the mainstream methodological choice for analyzing and interpreting histology images. In this paper, we present a comprehensive review of state-of-the-art deep learning approaches that have been used in the context of histopathological image analysis. From the survey of over 130 papers, we review the field's progress based on the methodological aspect of different machine learning strategies such as supervised, weakly supervised, unsupervised, transfer learning and various other sub-variants of these methods. We also provide an overview of deep learning based survival models that are applicable for disease-specific prognosis tasks. Finally, we summarize several existing open datasets and highlight critical challenges and limitations with current deep learning approaches, along with possible avenues for future research. 
\end{abstract}


\end{frontmatter}

%\linenumbers
%%%%%%%%%%%%%%%%%%%%%%%%%%%%%%%%%%%%%%%%%%%%%%%%%%%%%%%%%%%%%%%%%%%%%%%%%%%%%%%%%%%%%%%%%%%%%%%%%%%%%%%%%%%%
%% main text

\section{Introduction}
\label{sec:Introduction}
\vspace{-2mm}
The examination and interpretation of tissue sections stained with haematoxylin and eosin (H\&E) by anatomic pathologists is an essential component in the assessment of disease. In addition to providing diagnostic information, the phenotypic information contained in histology slides can be used for prognosis.  Features such as nuclear atypia, degree of gland formation, presence of mitosis and inflammation can all be indicative of how aggressive a tumour is, and may also allow predictions to be made about the likelihood of recurrence after surgery. Over the last 50 years, several scoring systems have been proposed that allow pathologists to grade tumours based on their appearance, for example, the Gleason score for prostate cancer \citep{epstein2005Gleason} and the Nottingham score for breast cancer \citep{Rakha2008Prognostic}. These systems provide important information to guide decisions about treatment and are valuable in assessing heterogeneous disease.  There is, however, considerable inter-pathologist variability, and some systems that require quantitative analysis, for example the residual cancer burden index \citep{Symmans2007Measurement}, are too time-consuming to use in a routine clinical setting. 

The first efforts to extract quantitative measures from microscopy images were in cytology. \cite{Prewitt1966} laid out the steps required for the “effective and efficient discrimination and interpretation of images” which described the basic paradigm of object detection, feature extraction and finally the training of a classification function that is still in use more than 50 years later. Early work in cytology and histopathology was usually limited to the analysis of the small fields of view that could be captured using conventional microscopy, and image acquisition was a time-consuming process \citep{Mukhopadhyay2018}. The introduction of whole slide scanners in the 1990s made it much easier to produce digitized images of whole tissue slides at microscopic resolution, and this led to renewed interest in the application of image analysis and machine learning techniques to histopathology. Many of the algorithms developed originally for computer-aided diagnosis in radiology have been successfully adapted for use in digital pathology, and \cite{Gurcan2009review, madabhushi2016image} provide comprehensive reviews of work carried out prior to the widespread adoption of deep learning methods.

In 2011, \cite{Beck2011stroma} demonstrated that features extracted from histology images could aid in the discovery of new biological aspects of cancer tissue, and \cite{Yuan2012tils} showed that features extracted from digital pathology images are complementary to genomic data. These advancements have led to a growing interest in the use of biomarkers extracted from digital pathology images for precision medicine \citep{bera2019artificial}, particularly in oncology. Later in 2012, \cite{Krizhevsky2012} showed that convolutional neural networks (CNNs) could outperform previous machine learning approaches by classifying 1.2 million high-resolution images in the ImageNet LSVRC-2010 contest into 1000 different classes. At the same time, \cite{NIPS2012_4741} showed that CNNs could outperform competing methods in segmenting nerves in electron microscopy images and detecting mitotic cells in histopathology images \citep{Cirecsan2013}. Since then, methods based on CNNs have consistently outperformed other handcrafted methods in a variety of deep learning (DL) tasks in digital pathology. The ability of CNNs to learn features directly from the raw data without the need for specialist input from pathologists and the availability of annotated histopathology datasets has also fueled the explosion of interest in deep learning applied to histopathology.

\begin{figure*}\centering
	\begin{minipage}{0.40\linewidth}
    \centerline{\includegraphics[width=\linewidth]{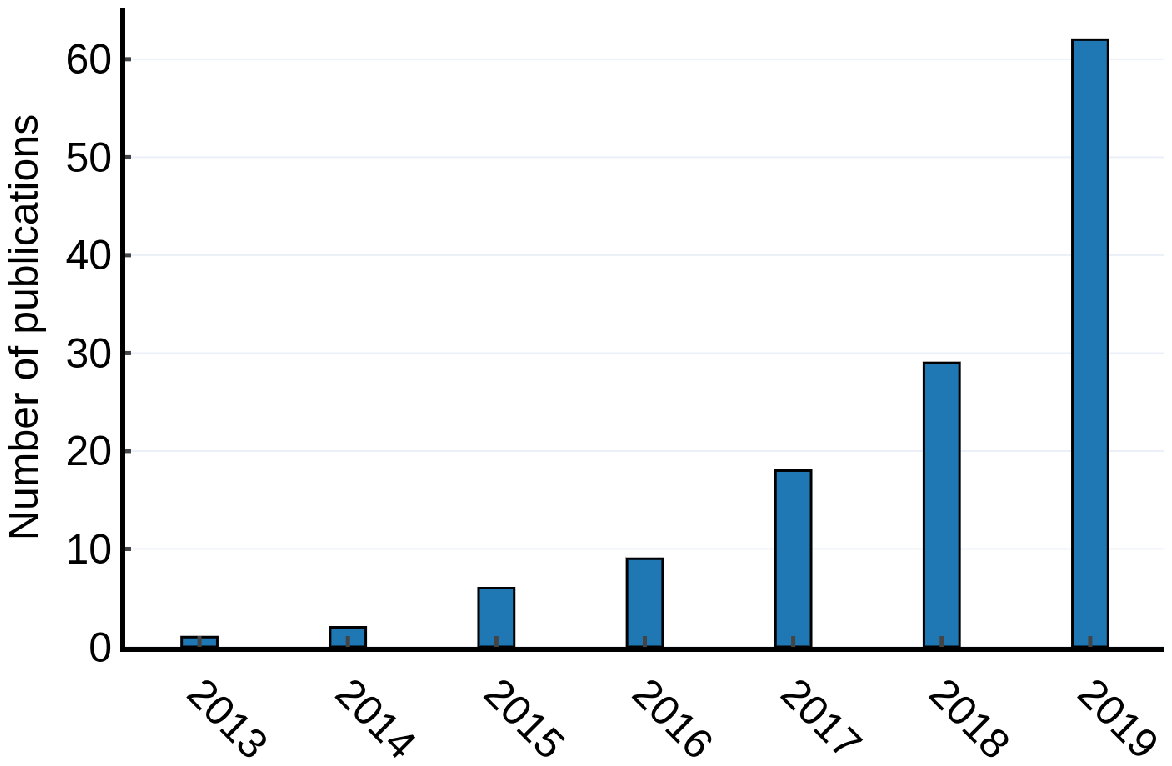}} 
    \centerline{(a)}\medskip
    \end{minipage}
\hfill
	\begin{minipage}{0.40\linewidth}
    \centerline{\includegraphics[width=\linewidth]{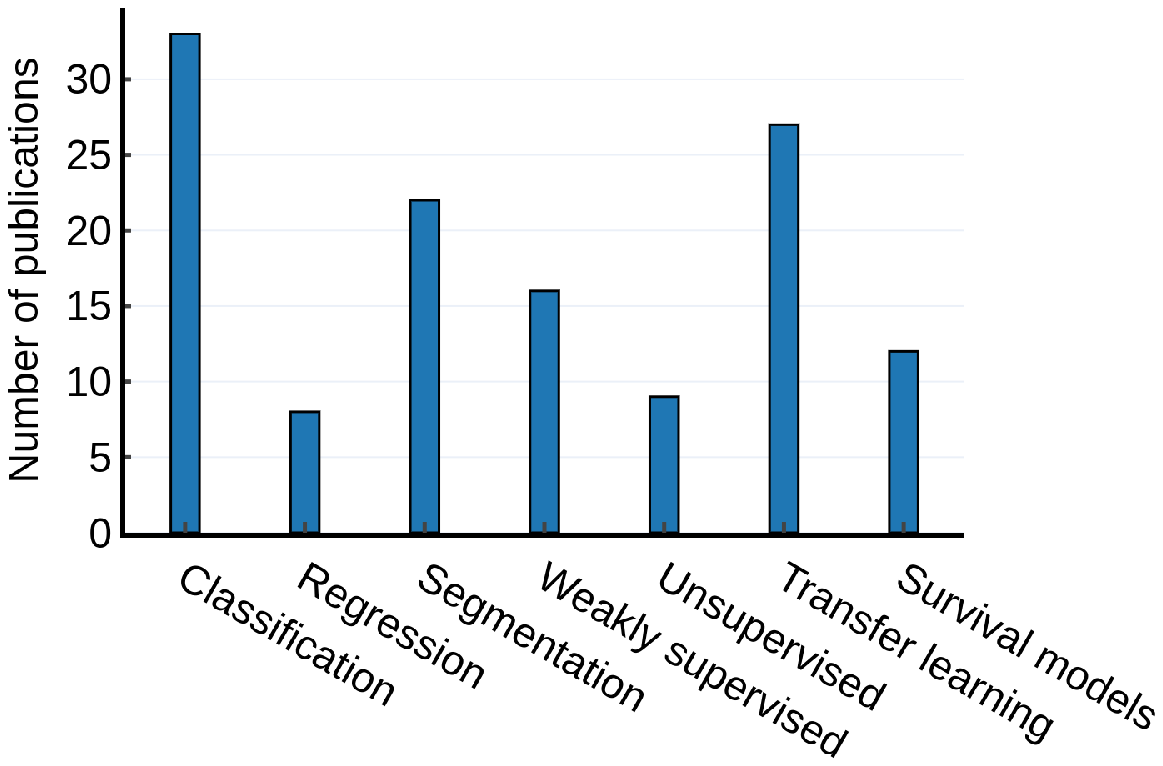}} 
    \vspace{-1.55mm}
    \centerline{(b)}\medskip
    \end{minipage}
    \vspace{-3.5mm}
\caption{(a) An overview of numbers of papers published from January 2013 to December 2019  in deep learning based computation histopathology surveyed in this paper. (b) A categorical breakdown of the number of papers published in each learning schemas.}
\label{Fig:Number of papers}
\end{figure*}

The analysis of whole-slide digital pathology images (WSIs) poses some unique challenges. The images are very large and have to be broken down into hundreds or thousands of smaller tiles before they can be processed. Both the context at low magnification, and the detail at high magnification, may be important for a task, therefore information from multiple scales needs to be integrated. In the case of survival prediction, salient regions of the image are not known \textit{a priori} and we may only have weak slide level labels. The variability within each disease subtype can be high and it usually requires a highly trained pathologist to make annotations. For cell based methods, many thousands of objects need to be detected and characterized. These challenges have made it necessary to adapt existing deep learning architectures and to design novel approaches specific to the digital pathology domain. In this work, we surveyed more than 130 papers, where deep learning has been applied to a wide variety of detection, diagnosis, prediction and prognosis tasks. We carried out this extensive review by searching  Google Scholar, PubMed and arXiv for papers containing keywords such as (``convolutional" or ``deep learning") and (``digital pathology" or ``histopathology" or ``computational pathology"). Additionally, we also included conference proceedings from MICCAI, ISBI, MIDL, SPIE and EMBC based on title/abstract of the papers. We also iterated over the selected papers to include any additional cross-referenced works that were missing from our initial search criteria. The body of research in this area is growing rapidly and this survey covers the period up to and including December 2019. A descriptive statistics of published papers according to their category and year is illustrated in Fig. \ref{Fig:Number of papers}.  
The remainder of this paper is organised as follows. Section \ref{sec:Overview of learning schemas} presents an overview of various learning schemes in DL literature in the context of computational histopathology. Section \ref{sec:Methodological approaches} discusses in detail different categories of DL schemes commonly used in this field. We categorize these learning mechanisms into supervised (Section \ref{ssec:Supervised learning}), weakly supervised (Section \ref{ssec:Weakly supervised learning}), unsupervised (Section \ref{ssec:Unsupervised methods}), transfer learning (Section \ref{ssec:Transfer learning}). Section \ref{sec:Survival models for disease prognosis} discusses survival models related to disease prognosis task. In Section \ref{sec:Discussion}, we discuss various open challenges including prospective applications and future trends in computational pathology, and finally, conclusions are presented in Section \ref{sec:Conclusions}.

\section{Overview of learning schemas}
\label{sec:Overview of learning schemas}

\begin{figure*}\centering
\centerline{\includegraphics[width=\linewidth]{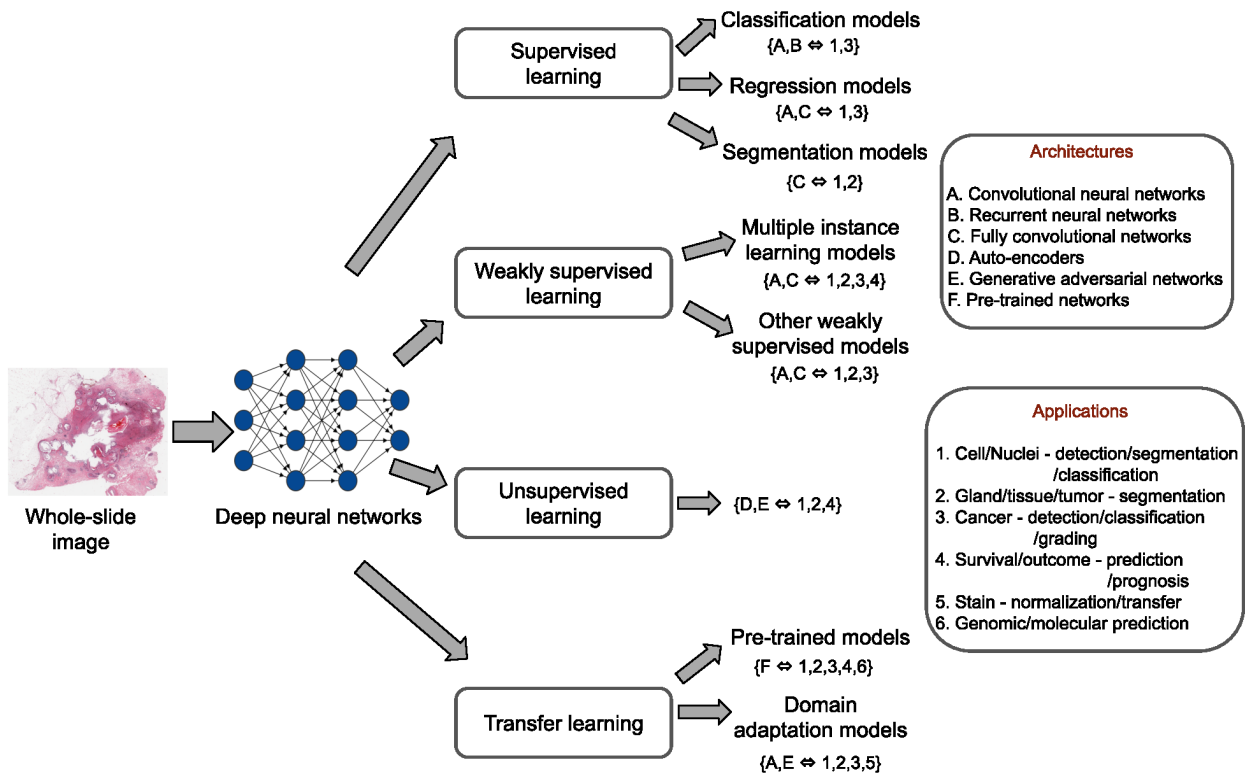}}
\caption{An overview of deep neural network models in computational histopathology. These models have been constructed using various deep learning architectures (shown in \textit{alphabetical} order) and applied to various histopathological image analysis tasks (depicted in \textit{numerical} order).}
\label{Fig:Overall_Learning_Scheme}
\end{figure*}
In this section, we provide a formal introduction to various learning schemes in the context of DL applied to computational pathology. These learning schemes are illustrated with an example of classifying a histology WSI as cancerous or normal. Based on these formulations, various DL models have been proposed in the literature, which are traditionally based on convolutional neural network (CNNs), recurrent neural networks (RNNs), generative adversarial networks (GANs), auto-encoders (AEs) and various other variants. For a detailed and thorough background of DL fundamentals and its existing architectures, we refer readers to \cite{Lecun2015deep, Goodfellow2016deep}, and with specific application of DL in medical image analysis to \cite{Litjens2017, Shen2017deep, Yi2019generative}.

In \textit{supervised learning}, we have a set of $N$ training examples $\{(x_i, y_i)\}_{i=1}^{N}$, where, each sample $x_{i} \in \mathbb{R} ^{C_h \times H \times W}$ is an input image (a WSI of dimension $H \times W$ pixels, with $C_h$ channels. For example, $C_h = 3$ channels for an RGB image) associated with a class label $y_{i} = \mathbb{R} ^{C}$, with $C$ possible classes. For example, in binary classification, $C$ takes the scalar form $\{0, 1\}$, and the set $\mathbb{R}$ for a regression task. The goal is to train a model $f_{\theta}:x \rightarrow y$ that best predicts the label for an unknown test image based on a loss function $\mathcal{L}$. For instance, $x$'s are the patches in WSIs and $y$'s are the labels annotated by the pathologist either as cancerous or normal. During the inference time, the model predicts the label of a patch from a previously unseen test set. This scheme is detailed in Section \ref{ssec:Supervised learning}, with an example illustrated in Fig. \ref{Fig:An overview of classification models.}. 

In \textit{weakly supervised learning} (WSL), the goal is to train a model $f_{\theta}$ using the readily available coarse-grained (image-level) annotations $C_i$, to automatically infer the fine-grained (pixel/patch)-level labels $c_i$. In histopathology, a pathologist labels a WSI as cancer, as long as a small part of this image contains cancerous region, without indicating its exact location. Such image-level annotations (often called \textit{``weak labels"}) are relatively easier to obtain in practice compared to expensive pixel-wise labels for supervised methods. An illustrative example for WSL scheme is shown in Fig. \ref{Fig:An overview of weakly supervised learning models.}, and this scheme is covered in-depth in Section \ref{ssec:Weakly supervised learning}. 

The \textit{unsupervised learning} aims at identifying patterns on the image, without mapping an input image sample into a predefined set of output (i.e. label). This type of models includes fully unsupervised methods, where the raw data comes in the form of images without any expert-annotated labels. A common technique in unsupervised learning is to transform the input data into a lower-dimensional subspace, and then group these lower-dimension representations (i.e. the latent vector) into mutually exclusive or hierarchical groups, based on a clustering technique. An example of unsupervised learning scheme is illustrated in Fig. \ref{Fig:An overview of unsupervised learning models.}, with existing methods in Section \ref{ssec:Unsupervised methods}.

In \textit{transfer learning} (TL), the goal is to transfer knowledge from one domain (i.e., source) to another domain (i.e., target), by relaxing the assumption that the train and test set must be independent and identically distributed. Formally, given a domain $\mathcal{D}=\{\mathcal{X},P(X)\}$, which is defined by the feature space $\mathcal{X}$, a marginal probability distribution $P(X)$ (where $X=\{x_1, \ldots, x_n\} \in \mathcal{X}$), and a task $\mathcal{T}=\{\mathcal{Y}, f(\cdot)\}$ - consisting of label space $\mathcal{Y}$ and a prediction function $f(\cdot)$. The aim of transfer learning is to improve the predictive function $f^\mathcal{T}(\cdot)$ in target domain ($\mathcal{D}^{t}$) by using the knowledge in source domain ($\mathcal{D}^{s}$) and source task ($\mathcal{T}^{s}$). For example, in histology, this scenario can occur while training a classifier on the source task $\mathcal{T}^{s}$ and possibly fine-tuning on a target task $\mathcal{T}^{t}$, with limited or no annotations. This scheme is explained in-detail in Section \ref{ssec:Transfer learning}. Note that, the \textit{domain adaptation}, which is a sub-field of transfer learning, is discussed thoroughly in Section \ref{sssec:Domain adaptation}. 

Next, we discuss various deep neural network (DNN) models in each of these learning schemes published in histopathology domain, along with the existing challenges and gaps in current research, and possible future directions in this perspective.

\section{Methodological approaches}
\label{sec:Methodological approaches}

The aim of this section is to provide a general reference guide to various deep learning models applied in computational histopathology from a methodological perspective. The DL models discussed in the following sections were originally developed for specific applications, but are applicable to a wide variety of histopathological tasks (Fig. \ref{Fig:Overall_Learning_Scheme}). Based on the learning schemes, the following sections are divided into supervised, weakly supervised, unsupervised and transfer learning approaches. The details are presented next.

\subsection{Supervised learning}
\label{ssec:Supervised learning}

\begin{figure*}\centering
\centerline{\includegraphics[width=0.90\linewidth]{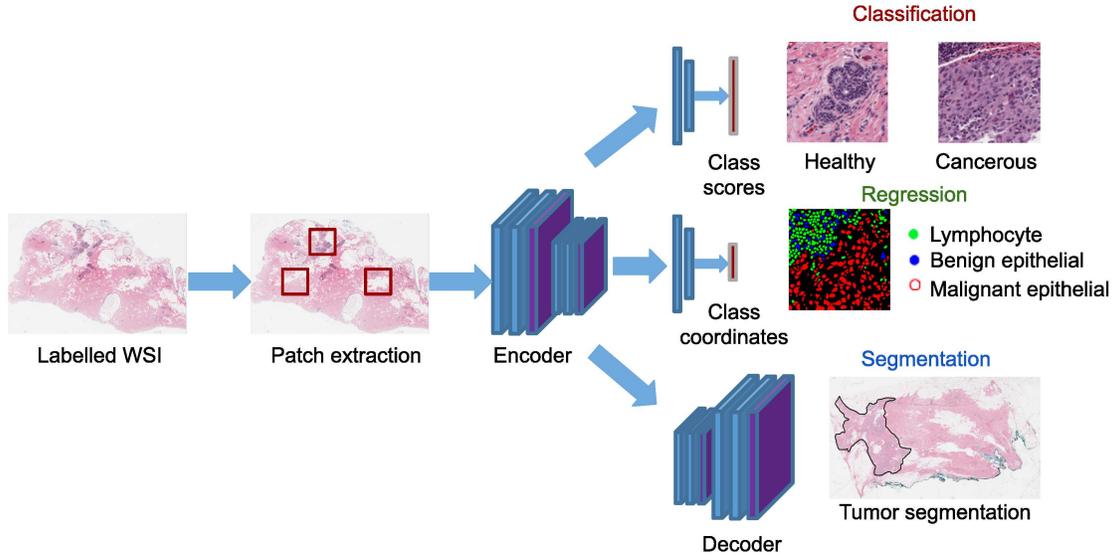}}
\caption{An overview of supervised learning models.}
\label{Fig:An overview of classification models.}
\end{figure*}
Among the supervised learning techniques, we identify three major canonical deep learning models based on the nature of tasks that are solved in digital histopathology: classification, regression and segmentation based models, as illustrated in Fig. \ref{Fig:An overview of classification models.}. The first category of models contains methods related to pixel-wise/sliding-window classification based approaches, which are traditionally formulated as object detection \citep{Girshick2015} or image classification tasks \citep{He2016} in the computer vision literature. The second category of models focuses on predicting the position of objects (e.g., cells or nuclei \citep{Sirinukunwattana2016}) or sometimes predicting a cancer severity score (e.g., H-score of breast cancer images \citep{Liu2019}) by enforcing topological/spatial constraints in DNN models. Finally, the last category of models is related to fully convolutional network (FCN) based approaches \citep{Long2015,Ronneberger2015}, which are widely adopted to solve semantic or instance segmentation problems in computer vision and medical imaging scenarios. The overview of papers in supervised learning is summarized in Table \ref{Tab:Overview of supervised learning models}. 

\subsubsection{Classification models}
\label{sssec:Classification models}

This category of methods uses a sliding window approach (i.e., patch centred on a pixel of interest) to identify objects (such as cells, glands, nuclei) or make image-level predictions (such as disease diagnosis and prognosis). Within this category, we further identify two sub-categories: (i) \textit{local-level tasks}, and (ii) \textit{global-level tasks}. The former stream of methods is based on a region (i.e., cell, nuclei) represented by a spatially pooled feature representations or scores, aiming at identifying or localizing objects. While the latter consists of methods related to image-level prediction tasks such as whole-slide level disease grading. 

\vspace{2mm}
\textit{A. Local-level task:}
Image classification such as detection of cells or nuclei is notably one of the most successful tasks, where deep learning techniques have made a tremendous contribution in the field of digital pathology. Methods based on CNNs have been extensively used for pixel-wise prediction task by a sliding window approach, to train the networks on small image patches rather than the entire WSI. Due to giga-resolution of WSIs (e.g., $100,000 \times 100,000$ pixels), applying a CNN directly to WSI is impractical, and hence, the entire WSI is divided into segments of small patches for analysis. In practice, these image patches are often annotated by the pathologist as a region containing an object of interest (e.g., cells/nuclei) or a background. A large corpus of deep learning methods applied to digital pathology is akin to computer vision models applied to visual object recognition task \citep{Russakovsky2015}.

The earliest seminal work proposed in \cite{Cirecsan2013} revolutionised the entire field of digital histopathology, by applying CNN based pixel prediction to detect mitosis in routinely stained H\&E breast cancer histology images. Subsequently, \cite{rao2018mitos} proposed a variant of the Faster-RCNN, which significantly outperformed all other competing techniques in both ICPR 2012 and the AMIDA 2013 mitosis detection challenge. The next set of methods were developed based on CNNs or a combination of CNN and handcrafted features. Since training of CNN models is often complex and requires a more extensive training set, the earliest works \citep{Wang2014, Kashif2016, Xing2016, Romo2016, Wang2016} focused on integrating CNN with biologically interpretable handcrafted features and these models showed excellent performance results in addressing the touching nuclei segmentation problem. A hybrid method, based on persistent homology, \cite{Qaiser2019} was able to capture the degree of spatial connectivity among touching nuclei, which is quite difficult to achieve using CNN models \citep{Sabour2017}.

Training a deep CNN from scratch requires large amounts of annotated data, which is very expensive and cumbersome to obtain in practice. A promising alternative is to use a pre-trained network (trained on a vast set of natural images, such as ImageNet) to fine-tune on a problem in different domain with limited number of annotations. Along these lines, \cite{Gao2017, Valkonen2019} proposed a fine-tuning based transfer learning approach, which consistently performed better than full training on a single dataset alone. In particular, \cite{Gao2017} made several interesting observations about improving CNN performance by optimising the hyperparameters of the network, augmenting the training data and fine-tuning rather than full training of the model. For more details and methods that are based on transfer learning are discussed thoroughly in Section \ref{ssec:Transfer learning}. 

Recent studies \citep{Albarqouni2016, irshad2017, amgad2019structured, marzahl2019fooling, hou2020dataset} have investigated the use of crowdsourcing approaches to alleviate the annotation burden on expert pathologists. Outsourcing labelling to non-experts can, however, lead to subjective and inconsistent labels and conventional DL models may find it challenging to train with noisy annotations. One way to improve the annotation quality is to collect multiple redundant labels per example and aggregate them via various voting techniques before training the model. For instance, \cite{Albarqouni2016} proposed to incorporate data aggregation as a part of the CNN learning process through an additional crowdsourcing layer for improving model performance. An alternative approach is to make use of expert advice (such as an experienced pathologist) by providing feedback for annotating rare and challenging cases \citep{amgad2019structured}. It is evident from the above studies that it is possible to train models using non-expert annotations successfully, but that care has to be taken to ensure quality. An easy and reliable way to obtain crowd labels for a large-scale database is to first obtain a set of precomputed annotations from an automated system, and correct only those labels with inconsistent markings under expert supervision \citep{marzahl2019fooling, hou2020dataset}. For example, \cite{marzahl2019fooling} showed the use of precomputed labels lead to an increase in model performance, which was independent of the annotator's expertise and that this reduced the interaction time by more than 30\% compared to other crowdsourcing approaches. A thorough and in-depth treatment of crowdsourcing methods applicable to medical imaging (including histopathology) is provided in \cite{orting2019survey}.

The addition of multi-scale and contextual knowledge into CNN plays an essential role in identifying overlapping cell structures in histopathology images. Conventional single scale models often suffer from two main limitations: 1) the raw-pixel intensity information around a small window does not have enough information about the degree of overlap between cells, and 2) use of a large window leads to an increase in the number of model parameters and training time. To alleviate these issues, several authors \citep{Song2015,Song2017} proposed a multi-scale CNN model to accurately solve the overlapping cell segmentation problem, with the addition of domain-specific shape priors during training. 
Despite several modifications to CNN architectures, however, traditional deep learning methods often lack generalisation ability due to stain variations across datasets and this is addressed in Section \ref{sssec:Stain normalization} 

In summary, among the bottom-up approaches, CNN is the current gold standard technique applied to a wide variety of low-level histopathology tasks such as cell or nuclei detection. Methods based on multi-scale CNN and transfer learning approaches are becoming increasingly popular due to their excellent generalization adaptability across a wide range of datasets and scanning protocols.   

\vspace{2mm}
\textit{B. Global-level task:}
Most of the published deep learning methods in this category focus on patch-based classification approach for whole-slide level disease prediction task. These techniques range from the use of simple CNN architectures \citep{Cruz-Rao2014, Ertosun2015} to more sophisticated models \citep{Qaiser2019b, Zhang2019} for accurate tissue-level cancer localization and WSI-level disease grading. For instance, \cite{Cruz-Rao2014, Cruz-Roa2017} proposed a simple 3-layer CNN for identifying invasive ductal carcinoma in breast cancer images which outperformed all the previous handcrafted methods by a margin of 5\%, in terms of average sensitivity and specificity. The main disadvantage of these methods is the relatively long computational time required to carry out a dense patch-wise prediction over an entire WSI. To address this issue, \cite{Cruz-Roa2018} proposed a combination of CNN and adaptive sampling based on quasi-Monte Carlo sampling and a gradient-based adaptive strategy, to precisely focus only on those regions with high-uncertainty. Subsequently, a few authors \citep{Litjens2016, Vandenberghe2017} employed a simpler patch-based CNN model for the  identification of breast and prostate cancer in WSI, achieving an AUC of 0.99 for the breast cancer experiment. In more recent years, some authors \citep{Bejnordi2018, Wei2019, Nagpal2019, Shaban2019, Halicek2019} have trained networks from scratch (i.e., full training) on huge set of WSIs. These networks include the most popular deep learning models traditionally used for natural image classification task such as \textit{VGGNet} \citep{Simonyan2014}, \textit{InceptionNet} \citep{Szegedy2015}, \textit{ResNet} \citep{He2016} and \textit{MobileNet} \citep{Howard2017} architectures. There is no generic rule about the choice of architectures, with the type of disease prediction task. However, the main success of these CNN models depends on the number of images available for training, choice of network hyper-parameters and various other boosting techniques \citep{Cirecsan2013, Nagpal2019} (Refer to Section \ref{sec:Critical analysis of architectures} for more details).

A few authors try to encode both local and global contextual information into CNN learning process for more accurate disease prediction in WSIs. Typically, contextual knowledge is incorporated into a CNN framework by modelling the spatial correlations between neighbouring patches, using the strengths of CNNs and conditional random fields (CRF) \citep{Zheng2015, Chen2017}. These techniques have been extensively used in computer vision tasks for sequence labeling \citep{Artieres2010, Peng2009} and semantic image segmentation problems \citep{Chen2017}. While in digital pathology, for instance, \cite{Kong2017} introduced a spatially structured network (Spatio-Net) combining CNN with 2D Long-short Term Memory (LSTM) to jointly learn the image appearance and spatial dependency features for breast cancer metastasis detection. A similar approach has also been adopted in \cite{Agarwalla2017} to aggregate features from neighbouring patches using 2D-LSTM's on WSIs. In contrast, \cite{Li2018} proposed an alternative technique based on CRF for modelling spatial correlations through a fully connected CRF component. The advantages of such models are that the whole DNN can be trained in an end-to-end manner with the standard backpropagation algorithm, with a slight overhead in complexity. Alternative methods have also been proposed to encode global contextual knowledge by adopting different patch level aggregation strategies. For example, \cite{Bejnordi2017b} employed a cascaded CNN model to aggregate patch-level pyramid representations to simultaneously encode multi-scale and contextual information for breast cancer multi-classification. Similarly, \cite{Awan2018} adopted a ResNet based patch classification model to output a high dimensional feature space. These features are then combined using a support vector machine (SVM) classifier to learn the context of a large patch, for discriminating different classes in breast cancer. 

Although the above methods include contextual information in the form of patch-based approaches, they still suffer from loss of visual context due to disjoint/random selection of small image patches. Furthermore, applying a CNN based classification model directly to WSI is computationally expensive, and it scales linearly with an increasing number of input image patches \citep{Qaiser2019b}. Some recent studies \citep{Qaiser2019b, Bentaieb2018, Xu2019} explored task-driven \textit{visual attention} models \citep{Mnih2014, Ranzato2014} for histopathology WSI analysis. Such models selectively focus on the most diagnostically useful areas (such as tissue components) while ignoring the irrelevant regions (such as the background) for further analysis. These kinds of visual attention models have been extensively explored in computer vision applications including object detection \citep{Liu2016fully}, image classification \citep{Mnih2014}, image captioning \citep{Sharma2015action}, and action recognition \citep{Xu2015show} tasks.

In routine clinical diagnosis, typically, a pathologist first examines different locations within a WSI to identify diagnostically indicative areas, and then combines this information over time across different eye fixations, to predict the presence or absence of cancer. This human visual attention mechanism can be modelled as a \textit{sequential learning} task in deep learning using RNNs. For instance, \cite{Qaiser2019b} modelled the prediction of immunohistochemical (IHC) scoring of HER2 \citep{Qaiser2018her} as a sequential learning problem, where the whole DNN is optimized via policy gradients trained under a deep reinforcement learning (DRL) framework. Furthermore, the authors also incorporated an additional task-specific mechanism to inhibit the model from revisiting the previously attended locations for further diagnosis. Similarly, \cite{Bentaieb2018, Xu2019} proposed recurrent attention mechanisms to selectively attend and classify the most discriminate regions in WSI for breast cancer prediction. Inspired by recent works \citep{Xu2015show, Krause2017hierarchical} in image captioning for natural scenes, \cite{Zhang2019} proposed an attention-based multi-modal DL framework to automatically generate clinical diagnostic descriptions and tissue localization attention maps, mimicking the pathologist. An attractive feature of their system is the ability to create natural language descriptions of the histopathology findings, whose structure closely resembles that of a standard clinical pathology report. 

In essence, attention-based models are gaining popularity in recent years and have several intriguing properties over traditional sliding-window (patch-based) approaches: i) by enforcing a region selection mechanism (i.e., attention), the model tries to learn only the most relevant diagnostically useful areas for disease prediction; ii) the number of model parameters is drastically reduced leading to faster inference time; and iii) the model complexity is independent of the size of WSI.

\subsubsection{Regression models}
\label{sssec:Regression models}

This category of methods focuses on detection or localization of objects by directly regressing the likelihood of a pixel being the centre of an object (e.g., cell or nucleus centre). Detection of cells or nuclei in histopathology images is challenging due to their highly irregular appearance and their tendency to occur as overlapping clumps, which results in difficulty in separating them as a single cell or a nucleus \citep{Naylor2018, Xie2018, Graham2019}. The use of pixel-based classification approaches for this task may result in suboptimal performance, as they do not necessarily consider the topological relationship between pixels that lie in the object centre with those in their neighbourhood \citep{Sirinukunwattana2016}. To tackle this issue, many authors cast the object detection task as a \textit{regression} problem, by enforcing topological constraints, such that the pixels near object centres have higher probability values than those further away. This formulation has shown to achieve better detection or localization of objects, even with significant variability in both the object appearance and their locations in images.

Deep regression models proposed in the literature are mainly based on either CNN or FCN architectures \citep{Long2015}. In the context of FCN, the earlier methods by \cite{Chen2016, Xie2018micro} proposed a simple FCN based regression model for detecting cells in histopathology images. The most recent methods attempt to improve the detection task by modifying the loss function \citep{Xie2018} or incorporating additional features into popular deep learning architectures \citep{Graham2019}. \cite{Xie2015,Xie2018} proposed a structured regression model based on fully residual convolutional networks for detecting cells. The authors adopted a weighted MSE loss by assigning higher weights to misclassified pixels that are closer to cell centres. A similar approach by \cite{Xing2019}, adopted a residual learning based FCN architecture for simultaneous nucleus detection and classification in pancreatic neuroendocrine tumour Ki-67 images. In their model, an additional auxiliary task (i.e., ROI extraction) is also introduced to assist and boost the nucleus classification task using weak annotations. To solve the challenging touching nuclei segmentation problem, \cite{Naylor2018} proposed a model to identify superior markers for the watershed algorithm by regressing the intra-nuclear distance map. \cite{Graham2019} went one step further, proposing a unified FCN model for simultaneous nuclear instance segmentation and classification which effectively encodes both the horizontal and vertical distance information of nuclei pixels to their centre of mass for accurate nuclei separation in multi-tissue histology images.

Other authors adopted alternative methods by modifying the output layer of CNN, to include distance constraints or a voting mechanism into the network learning process. For instance, \cite{Sirinukunwattana2016} introduced a new layer modifying the output of a CNN to predict a probability map which is topologically constrained, such that the high confidence scores are likely to be assigned to the pixels closer to nuclei centre in colon histology images. This method was later extended in \cite{Swiderska2019} to detect lymphocytes in immunohistochemistry images. \cite{Xie2015} proposed an alternative method based on the voting mechanism for nuclei localization. This can be viewed as an implicit Hough-voting codebook, which learns to map an image patch to a set of voting offsets (i.e., nuclei positions) and the corresponding confidence scores to weight each vote. This set of weighted votes is then aggregated to estimate the final density map used to localize the nuclei positions in neuroendocrine tumour images.

\subsubsection{Segmentation models}
\label{sssec:Segmentation models}

Segmentation of histological primitives such as cells, glands, nuclei and other tissue components is an essential pre-requisite for obtaining reliable morphological measurements to assess the malignancy of several carcinomas \citep{Chen2017dcan,Sirinukunwattana2017,Bulten2019automated}. Accurate segmentation of structures from histology images often requires the pixel-level delineation of object contour or the whole interior of the object of interest. CNNs trained to classify each patch centred on a pixel of interest as either foreground or background, can be used for segmentation tasks by employing a sliding-window approach. However, given the large size of giga-pixel WSIs, patch-based approaches lead to a large number of redundant computations in overlapping regions, in turn resulting in a drastic increase in computational complexity and loss of contextual information \citep{Chen2017dcan,Lin2019fast}. The other alternative is to employ fully convolutional networks (FCN) \citep{Long2015, Ronneberger2015}, which take as input an arbitrary sized image (or a patch) and output a similar-sized image in a single forward pass. The whole FCN model can be trained via end-to-end backpropagation and directly outputs a dense per-pixel prediction score map. Hence, segmentation models in histopathology are mainly built on the representative power of FCN and its variants, which are generally formulated as a \textit{semantic segmentation} task, with applications ranging from nucleus/gland/duct segmentation \citep{Kumar2019,Sirinukunwattana2017,seth2018automated} to the prediction of cancer \citep{Liu2019,Bulten2019epithelium} in WSIs. 

In order to determine an optimal model suitable for a given task, \cite{Swiderska2019, De2018} compared FCN with UNet architecture \citep{Ronneberger2015} and found that better generalization ability and robustness was achieved using a UNet model. The key feature of the UNet is the upsampling path of the network, which learns to propagate the contextual information to high-resolution layers, along with additional skip connections to yield more biologically plausible segmentation maps, compared to the standard FCN model. The traditional FCN model also lacks smoothness constraints, which can result in poor delineation of object contours and formation of spurious regions while segmenting touching/overlapping objects \citep{Zheng2015}. To circumvent this problem, \cite{Bentaieb2016} formulated a new loss function to incorporate boundary smoothness and topological priors into FCN learning, for discriminating epithelial glands with other tissue structures in histology images. 

The appearance of histological objects such as glands and nuclei vary significantly in their size, shape and often occur as overlapping clumped instances, which makes them difficult to distinguish with the other surrounding structures. A few methods attempted to address this issue by leveraging the representation power of FCN with multi-scale feature learning strategies \citep{Chen2017,Lin2017feature}; to effectively delineate varying size objects in histology images. For instance, \cite{Chen2017dcan} proposed a multi-level contextual FCN with auxiliary supervision mechanism \citep{Xie2015holistically} to segment both glands and nuclei in histology images. They also devised an elegant multi-task framework to integrate object appearance with contour information, for precise identification of touching glands. This work was later extended in \cite{Van2018segmentation} by combining the efficient techniques of DCAN \citep{Chen2017dcan}, UNet, and identity mapping in ResNet to build an FCN model for segmenting epithelial glands in double-stained images. 

Some authors have proposed variants of FCN to enhance segmentation - in particular at glandular boundaries, by compensating for the loss occurring in max-pooling layers of FCNs. For example, \cite{Graham2019mild} introduced minimum information loss dilated units in residual FCNs, to help retain the maximal spatial resolution critical for segmenting glandular structures at boundary locations. Later, \cite{Ding2019} employed a similar technique to circumvent the loss of global information by introducing a high-resolution auxiliary branch in the multi-scale FCN model, to locate and shape the glandular objects. \cite{Zhao2019} proposed a feature pyramid based model \citep{Lin2017feature} to aggregate local-to-global features in FCN, to enhance the discriminative capability of the model in identifying breast cancer metastasis. Moreover, they also devised a synergistic learning approach to collaboratively train both the primary detector and an extra decoder with semantic guidance, to help improve the model's ability to retrieve metastasis.

Conventional FCN based models are fundamentally designed to predict the class label for each pixel as either foreground or background, but are unable to predict the individual object instances (i.e., recognizing the categorical label of foreground pixels). In computer vision, such problems can be formulated as an \textit{``instance-aware semantic segmentation"} task \citep{Hariharan2014simultaneous, Li2017fully}, where segmentation and classification of object instances are performed simultaneously in a joint end-to-end manner. In histology, \cite{Xu2017} formulated the gland instance segmentation as two sub-tasks - gland segmentation and instance recognition task, using a multi-channel deep network model \citep{Dai2016instance}. The gland segmentation is performed using FCN, while, the gland instance boundaries are recognized using the location \citep{Girshick2015} and boundary cues \citep{Xie2015holistically}. A similar formulation has been adopted in \cite{Qu2019} to solve the joint segmentation and classification of nuclei using an FCN trained with perceptual loss \citep{Johnson2016perceptual}.  

Most deep learning methods in digital pathology are applied on small-sized image patches rather than the entire WSI, restricting the prediction ability of the model to a narrow field-of-view. The conventional patch-based approaches often suffer from three main limitations: i) the extracted individual patches from WSI have a narrow field-of-view, with limited contextual knowledge about the surrounding structures; ii) patch-based models are not consistent with the way a pathologist analyzes a slide under a microscope; and iii) a large number of redundant computations are carried out in overlapping regions, resulting in increased computational complexity and slower inference speed. In order to alleviate the first two issues, attempts have been made to mimic the way in which a pathologist usually analyzes a slide at various magnification levels before arriving at the final decision. Such mechanisms are integrated into the FCN model by designing multi-magnification networks \citep{Ho2019, Tokunaga2019}, each trained on different field-of-view image patches to obtain a better discriminative feature representation compared to a single-magnification model. For instance, \cite{Ho2019} proposed a multi-encoder and multi-decoder FCN model utilizing multiple input patches at various magnification levels (e.g., 20x, 10x and 5x) to obtain intermediate feature representations that are shared among each FCN model for accurate breast cancer image segmentation. A similar approach has been adopted in \cite{Tokunaga2019, Gecer2018} by training multiple FCN's on different field-of-view images, which are aggregated to obtain a final segmentation map. In contrast, \cite{Gu2018} designed a multiple encoder model to aggregate information across different magnification levels, but utilized only one decoder to generate a final prediction map. 

Nevertheless, the above patch-based models still suffer from significant computational overhead at higher magnification levels, and hence, do not scale well to WSIs. Therefore, some authors \citep{Lin2019fast,Lin2018scannet} have proposed a variant of FCN which consists of a dense scanning mechanism, that shares computations in overlapping regions during image scanning. To further improve the prediction accuracy of the FCN model, a new pooling layer named as `anchor layer' is also introduced in \cite{Lin2019fast} by reconstructing the loss occurred in max-pooling layers. Such models have been shown to have inference speeds a hundred times faster than traditional patch-based approaches, while still ensuring a higher prediction accuracy in WSI analysis. On the other hand, \cite{Guo2019fast} presented an alternative method for fast breast tumour segmentation, in which, a network first pre-selects the possible tumour area via CNN based classification, and later refines this initial segmentation using an FCN based model. Their proposed framework obtains dense predictions with 1/8 size of original WSI in 11.5 minutes (on CAMELYON16 dataset), compared to the model trained using FCN alone.

%%%%%%%%%%%%%%%%%%%%%%%%%%%%%%%%%%%%%%%%%%%%%%%%%%%%%%%%%%%%%%%%%%%%%%%%%%%%%%%%%%%%%%%%%%%%%%%%%%
\vfill\null\vspace*{-100cm}
%%%%%%%%%%%%%%%%%%%%%%%%%%%%%%%%%%%%%%%%%% LONG Table %%%%%%%%%%%%%%%%%%%%%%%%%%%%%%%%%%%%%%%%%%%%%%%%
\begin{center}
\onecolumn
\scriptsize
\begin{longtable}{p{3.5cm}p{1.3cm}p{1cm}p{2.75cm}p{4cm}p{3.5cm}}
\caption{Overview of supervised learning models. The acronyms for the staining stands for: H\&E (haematoxylin and eosin); DAB-H (Diaminobenzidine-Hematoxylin); IFL (Immunofluorescent); ER (Estrogen receptor), PR (Progesterone receptor); PC (Phase contrast); HPF (High power field); Pap (Papanicolaou stain); PHH3 (Phosphohistone-H3); IHC (Immunohistochemistry staining); PAS (Periodic acid–Schiff). Note: (\checkmark) indicates the code is publicly available and the link is provided in their respective paper.}
\label{Tab:Overview of supervised learning models} \\
\toprule[0.75pt]
Reference & \multicolumn{1}{l}{Cancer types} & \multicolumn{1}{l}{Staining} & \multicolumn{1}{l}{Application} & \multicolumn{1}{l}{Method} & \multicolumn{1}{l}{Dataset} \\ \midrule 
\multicolumn{6}{l}{Classification models} \\ \hline \noalign{\vskip 1mm}
\multicolumn{6}{l}{\textit{A. Local-level task}} \\ \hline \noalign{\vskip 1mm}
% 2013
\cite{Cirecsan2013} & Breast & H\&E & Mitosis detection & Pixel based CNN classifier & ICPR2012 (50 images)\\
\cite{Wang2014} & Breast & H\&E & Mitosis detection & Cascaded ensemble of CNN + handcrafted features & ICPR2012 (50 images) \\
\cite{Song2015} & Cervix & H\&E & Segmentation of cervical cytoplasm and nuclei & Multi-scale CNN + graph-partitioning approach & Private set containing 53 cervical cancer images \\
\cite{Kashif2016} & Colon & H\&E & Cell detection & Spatially constrained CNN + hand-crafted features & 15 images of colorectal cancer tissue images\\
\cite{Xing2016} & Multi-Cancers & H\&E, IHC & Nuclei segmentation & CNN + selection-based sparse shape model & Private set containing brain tumour (31), pancreatic NET (22), breast cancer (35) images \\
\cite{Romo2016} & Breast & H\&E & Tubule nuclei detection and classification & CNN based classification of pre-detected candidate nuclei & 174 images with ER(+) breast cancer cases \\
\cite{Wang2016} & Lung & H\&E & Cell detection & Two shared-weighted CNNs for joint cell detection and classification & TCGA (300 images) \\
\cite{Albarqouni2016} & Breast & H\&E & Mitosis detection & Multi-scale CNN via crowdsourcing layer & AMIDA2013 (666 - HPF images)\\
% 2017
\cite{Song2017} & Cervix & Pap, H\&E & Segmentation of cervical cells & Multi-scale CNN model & Overlapping cervical cytology image segmentation challenge (ISBI 2015) - 8 images, private set - 21 images \\
\cite{Gao2017} & Multi-Cancers & IFL & Cell classification & CNN (LeNet-5) based classification of HEp2-cells & ICPR2012 (28 images), ICPR2014 (83 images) \\
%2018
\cite{rao2018mitos} & Breast & H\&E & Mitosis detection & Faster RCNN based multi-scale region proposal model & ICPR2012 (50 images), AMIDA2013 (23 images), ICPR2014 (2112 images)\\
\cite{Tellez2018} & Breast & H\&E, PHH3 & Mitosis detection & Ensemble of CNNs using H\&E registered to PHH3 tissue slides as reference standard & TNBC (36 images), TUPAC (814 images) \\
% 2019
\cite{Qaiser2019} & Colon & H\&E & tumour segmentation & Combination of CNN and persistent homology feature based patch classifier & Two private sets containing 75 and 50 colorectal adenocarcinoma WSIs \\
\hline \noalign{\vskip 1mm}
\multicolumn{6}{l}{\textit{B. Global-level task}} \\ \hline \noalign{\vskip 1mm} 
\cite{Cruz-Rao2014} & Breast & H\&E & Detection of invasive ductal carcinoma & CNN based patch classifier & Private set - 162 cases \\
\cite{Ertosun2015} & Brain & H\&E & Glioma grading & Ensemble of CNN models & TCGA (54 WSIs)\\
\cite{Litjens2016} & Multi-Cancers & H\&E & Detection of prostate and breast cancer & CNN based pixel classifier & Two private sets (225 + 173 WSIs) \\
\cite{Bejnordi2017b} & Breast & H\&E & Breast cancer classification & Stacked CNN incorporating contextual information & Private set - 221 images \\
\cite{Agarwalla2017} & Breast & H\&E & tumour segmentation & CNN + 2D-LSTM for representation learning and context aggregation & Camelyon16 (400 WSIs) \\
\cite{Kong2017} & Breast & H\&E & Detection of breast cancer metastases & CNN with the 2D-LSTM to learn spatial dependencies between neighboring patches & Camelyon16 (400 WSIs) \\
\cite{Vandenberghe2017} & Breast & IHC & IHC scoring of HER2 status in breast cancer & CNN based patch classifier & 71 WSIs of invasive breast carcinoma (Private set) \\
\cite{Cruz-Roa2017} & Breast & H\&E & Detection of invasive breast cancer & CNN based patch classifier & TCGA + four other private sets (584 cases)\\
\cite{sharma2017deep} & Stomach & H\&E, IHC & Gastric cancer classification and necrosis detection & Patch-based CNN classifier & Private set - 454 WSIs \\
% 2018
\cite{Bentaieb2018} (\checkmark) & Breast & H\&E & Detection of breast cancer metastases & CNN based recurrent visual attention model & Camelyon16 (400 WSIs)\\
\cite{Awan2018} & Breast & H\&E & Breast cancer classification & CNN based patch classification model incorporating contextual information & BACH 2018 challenge (400 WSIs) \\
\cite{Li2018} (\checkmark) & Breast & H\&E & Detection of breast cancer metastases & CNN + CRF to model spatial correlations between neighboring patches & Camelyon16 (400 WSIs) \\
\cite{Bejnordi2018} & Breast & H\&E & Detection of invasive breast cancer & Multi-stage CNN that first identifies tumour-associated stromal alterations and further classify into normal/benign vs invasive breast cancer & Private set - 2387 WSIs \\
\cite{Cruz-Roa2018} & Breast & H\&E & Detection of invasive breast cancer & Patch based CNN model with adaptive sampling method to focus only on high uncertainty regions & TCGA + 3 other public datasets (596 cases) \\
%2019
\cite{Qaiser2019} & Breast & IHC & Immunohistochemical scoring of HER2 & Deep reinforcement learning model that treats IHC scoring as a sequential learning task using CNN + RNN & HER2 scoring contest (172 images), private set - 82 gastroenteropancreatic NET images \\ 
\cite{Wei2019} (\checkmark) & Lung & H\&E & Classifcation of histologic subtypes on lung adenocarcinoma & ResNet-18 based patch classifier & Private set - 143 WSIs \\ 
\cite{Nagpal2019} & Prostate & H\&E & Predicting Gleason score & CNN based regional Gleason pattern classification + k-nearest-neighbor based Gleason grade prediction & TCGA (397 cases) + two private sets (361 + 11 cases) \\
\cite{Shaban2019} (\checkmark) & Mouth & H\&E & tumour infiltrating lymphocytes abundance score prediction for disease free survival & CNN (MobileNet) based patch classifier, followed by statistical analysis & 70 cases of oral squamous cell carcinoma WSIs (Private set)\\
\cite{Halicek2019} & Head \& Neck & H\&E & Detection of squamous cell carcinoma and thyroid carcinoma & CNN (Inception-v4) based patch classifier & Private set - 381 images \\  
\cite{Xu2019} & Breast & H\&E & Detection of breast cancer & Deep hybrid attention (CNN + LSTM) network & BreakHis (7,909 images) \\
\cite{Zhang2019} (\checkmark) & Bladder & H\&E & Bladder cancer diagnosis & CNN + RNN to generate clinical diagnostic descriptions and network visual attention maps & 913 images of urothelial carcinoma from TCGA and private set\\
\\ \hline \noalign{\vskip 1mm}
\multicolumn{6}{l}{Regression models} \\ \hline \noalign{\vskip 1mm}
% 2015
\cite{Xie2015} & Multi-Cancers & Ki-67 & Nuclei detection & CNN based hough voting approach & Neuroendocrine tumour set (private - 44 images)\\ 
\cite{Xie2015beyond} & Multi-Cancers & H\&E, Ki-67 & Cell detection & CNN based structured regression model & TCGA (Breast-32 images), HeLa cervical cancer (22 images), Neuroendocrine tumour images (60 images)\\ 
% 2016
\cite{Chen2016} & Breast & H\&E & Mitosis detection & FCN based deep regression network & ICPR2012 (50 images)\\
\cite{Sirinukunwattana2016} & Colon & H\&E & Nuclei detection and classification & CNN with spatially constrained regression & CRCHisto (100 images)\\
% 2018
\cite{Naylor2018} (\checkmark) & Multi-Cancers & H\&E & Nuclei segmentation & CNN based regression model for touching nuclei segmentation & TNBC (50 images), MoNuSeg (30 images)\\
\cite{Xie2018} & Multi-Cancers & H\&E, Ki-67 & Cell detection & Structured regression model based on fully residual CNN & TCGA (Breast-70 image patches), Bone marrow (11 image patches), HeLa cervical cancer (22 images), Neuroendocrine tumour set (59 image patches)\\
% 2019
\cite{Graham2019} (\checkmark) & Multi-Cancers & H\&E & Nuclei segmentation and classification & CNN based instance segmentation and classification framework & CoNSeP (41 images), MoNuSeg (30 images), TNBC (50 images), CRCHisto (100 images), CPM-15 (15 images), CPM-17 (32 images)\\
\cite{Xing2019} & Pancreas & Ki-67 & Nuclei detection and classification & FCN based structured regression model & Pancreatic neuroendocrine tumour set (private - 38 images)
\\ \hline \noalign{\vskip 1mm}
\multicolumn{6}{l}{Segmentation models} \\ \hline \noalign{\vskip 1mm}
%2016
\cite{Bentaieb2016} & Colon & H\&E & Segmentation of colon glands & A loss function accounting for boundary smoothness and topological priors in FCN learning & GLAS challenge (165 images)\\
%2017
\cite{Chen2017dcan} & Multi-Cancers & H\&E & Segmentation of glands and nuclei & Multi-task learning framework with contour-aware FCN model for instance segmentation & GLAS challenge (165 images), MICCAI 2015 nucleus segmentation challenge (33 images) \\
\cite{Xu2017} & Colon & H\&E & Segmentation of colon glands & Multi-channel deep network model for gland segmentation and instance recognition & GLAS challenge (165 images) \\ 
%2018
\cite{De2018} & Kidney & PAS & Segmentation of renal tissue structures & Evaluated three different architectures: FCN, Multi-scale FCN and UNet & 15 WSIs of renal allograft resections (private set) \\
\cite{Van2018segmentation} & Colon & H\&E, IHC & Segmentation of glandular epithelium in H\&E and IHC staining images & CNN model based on integration of DCAN, UNet and ResNet models & GLAS challenge (165 images) and a private set containing colorectal tissue microarray images \\
\cite{Gecer2018} & Breast & H\&E & Detection and classification of breast cancer & Ensemble of multi-scale FCN's followed by CNN based patch classifier & 240 breast histopathology WSIs (private set)\\
\cite{Gu2018} & Breast & H\&E & Detection of breast cancer metastasis & UNet based multi-resolution network with multi-encoder and single decoder model & Camelyon16 (400 images)\\ 
\cite{Guo2019fast} & Breast & H\&E & Detection of breast cancer metastasis & Classification (Inception-V3) based semantic segmentation model (DCNN) & Camelyon16 (400 images) \\
%2019
\cite{Bulten2019automated} & Prostate & H\&E & Grading of prostate cancer & UNet based segmentation of Gleason growth patterns, followed by subsequent cancer grading & 1243 WSIs of prostate biopsies (private set) \\
\cite{Lin2019fast} & Breast & H\&E & Detection of breast cancer metastasis & FCN based model for fast inference of WSI analysis & Camelyon16 (400 WSIs) \\
\cite{Liu2019} & Breast & DAB-H & Immunohistochemical scoring for breast cancer & Multi-stage FCN framework that directly predicts H-Scores of breast cancer TMA images & 105 TMA images of breast adenocarcinomas (private set) \\
\cite{Bulten2019epithelium} & Prostate & IHC, H\&E & Segmentation of epithelial tissue & Pre-trained UNet on IHC is used as a reference standard to segment epithelial structures in H\&E WSIs & 102 prostatectomy WSIs \\
\cite{Swiderska2019} & Multi-Cancers & IHC & Lymphocyte detection & Investigated the effectiveness of four DL methods - FCN, UNet, YOLO and LSM & LYON19 (test set containing 441 region-of-interests (ROIs)) \\
\cite{Graham2019mild} & Colon & H\&E & Segmentation of colon glands & FCN with minimum information loss units and atrous spatial pyramid pooling & GLAS challenge (165 images), CRAG dataset (213 images)\\
\cite{Ding2019} & Colon & H\&E & Segmentation of colon glands & Multi-scale FCN model with a high-resolution branch to circumvent the loss in max-pooling layers & GLAS challenge (165 images), CRAG dataset (213 images) \\
\cite{Zhao2019} & Breast & H\&E & Detection and classification of breast cancer metastasis & Feature pyramid aggregation based FCN network with synergistic learning approach & Camelyon16 (400 WSIs), Camelyon17 (1000 WSIs) \\
\cite{Qu2019} (\checkmark) & Lung & H\&E & Nuclei segmentation and classification & FCN trained with perceptual loss & 40 tissue images of lung adenocarcinoma (private set) \\
\cite{Ho2019} & Breast & H\&E & Breast cancer multi-class tissue segmentation & Deep multi-magnification model with multi-encoder, multi-decoder and multi-concatenation network & Private set containing TNBC (38 images) and breast margin dataset (10 images)\\
\cite{Tokunaga2019} & Lung & H\&E & Segmentation of multiple cancer subtype regions & Multiple UNets trained with different FOV images + an adaptive weighting CNN for output aggregation & 29 WSIs of lung adenocarcinoma (private set)\\
\cite{Lin2019fast} & Breast & H\&E & Detection of breast cancer metastasis & FCN based model with anchor layers for fast and accurate prediction of cancer metastasis & Camelyon16 (400 images)\\
\cite{Pinckaers2019neural} (\checkmark) & Colon & H\&E & Segmentation of colon glands & Incorporating neural ordinary differential equations in UNet to allow an adaptive receptive field & GLAS challenge (165 images)\\
\cite{seth2018automated} & Breast & H\&E & Segmentation of DCIS & Compared UNets trained at multiple resolutions &  training:183 WSIs, testing:19 WSIs (private set) \\ 
\bottomrule [0.75pt]
\end{longtable}
\twocolumn
\end{center}
%%%%%%%%%%%%%%%%%%%%%%%%%%%%%%%%%%%%%%%%%%%%%%%%%%%%%%%%%%%%%%%%%%%%%%%%%%%%%%%%%%%%%%%%%%%%%%%%%%
% \vfill\null\vspace*{-25cm}
%%%%%%%%%%%%%%%%%%%%%%%%%%%%%%%%%%%%%%%%%%%%%%%%%%%%%%%%%%%%%%%%%%%%%%%%%%%%%%%%%%%%%%%%%%%%%%%%%%

\subsection{Weakly supervised learning}
\label{ssec:Weakly supervised learning}
\begin{figure*}\centering
\centerline{\includegraphics[width=\linewidth]{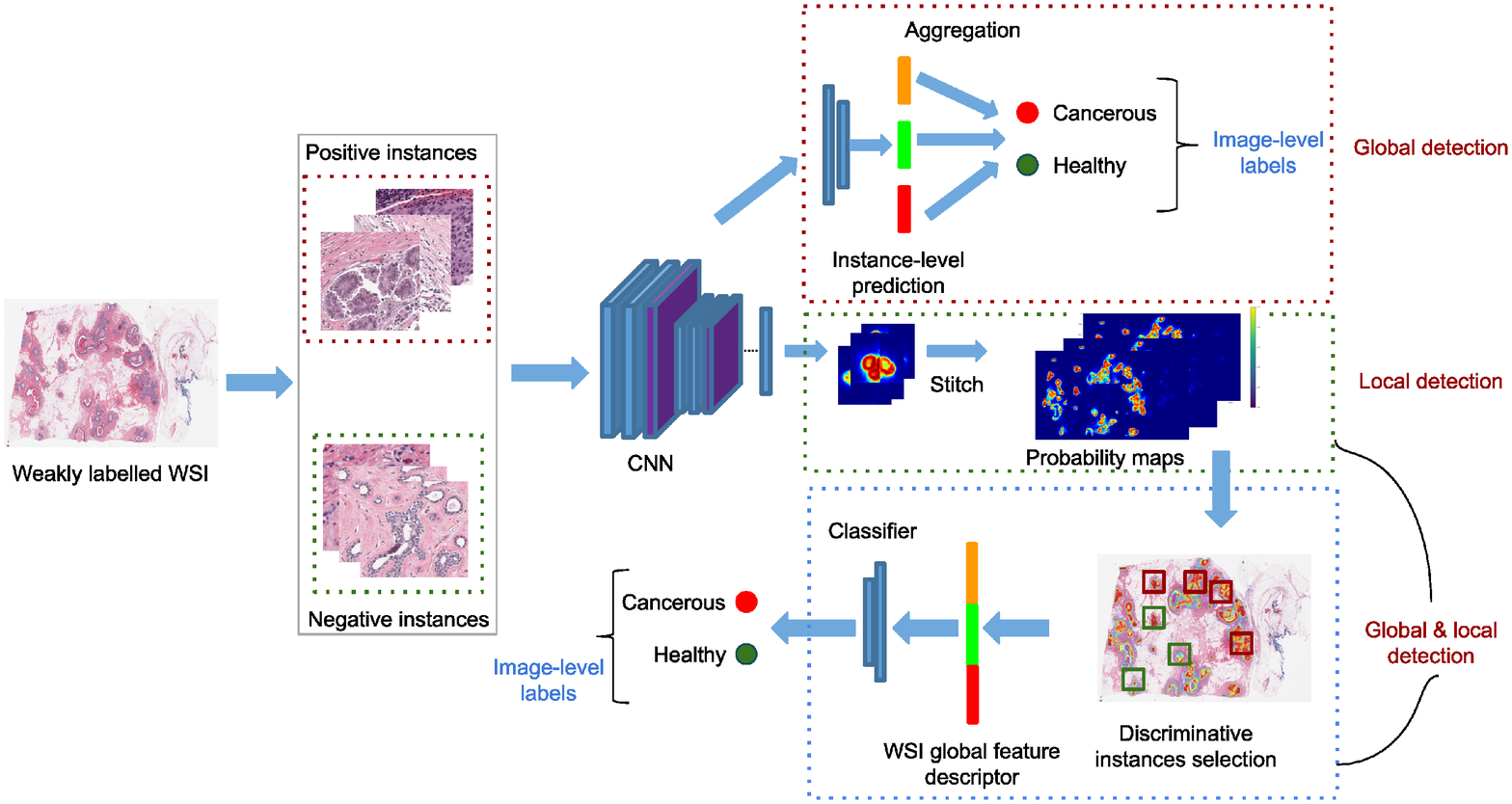}}
\caption{An overview of weakly supervised learning models.}
\label{Fig:An overview of weakly supervised learning models.}
\end{figure*}
%%%%%%%%%%%%%%%%%%%%%%%%%%%%%%%%%%%%%%%%%%%%%% Table %%%%%%%%%%%%%%%%%%%%%%%%%%%%%%%%%%%%%%%%%%%%%%%%%%%%%%%%%%%%%%%
\begin{table*}
\scriptsize
\centering
\caption{Overview of weakly supervised learning models. Note: (\checkmark) indicates the code is publicly available and the link is provided in their respective paper.}
\label{Tab:Overview of weakly supervised learning models}
\begin{tabularx}{\textwidth}{lllXXX}
\toprule[0.75pt]
Reference & \multicolumn{1}{l}{Cancer types} & \multicolumn{1}{l}{Staining} & \multicolumn{1}{l}{Application} & \multicolumn{1}{l}{Method} & \multicolumn{1}{l}{Dataset} \\ \midrule 
\multicolumn{6}{l}{Multiple instance learning (MIL)} \\ \hline \noalign{\vskip 1mm}
\cite{Hou2015efficient} & Brain & H\&E & Glioma subtype
classification & Expectation-maximization based MIL with CNN + logistic regression & TCGA (1,064 slides) \\
\cite{Jia2017} & Colon & H\&E & Segmentation of cancerous regions & FCN based MIL + deep supervision and area constraints & Two private sets containing colon cancer images (910+60 images) \\
\cite{Liang2018weakly} & Stomach & H\&E & Gastric tumour segmentation & Patch-based FCN + iterative learning approach & China Big Data and AI challenge (1,900 images)\\
\cite{Ilse18} (\checkmark) & Multi-Cancers & H\&E & Cancer image classification & MIL pooling based on gated-attention mechanism & CRCHisto (100 images)\\
\cite{Wang2019rmdl} & Stomach & H\&E & Gastric cancer detection & Two-stage CNN framework for localization and classification & Private set (608 images) \\
\cite{Wang2019} & Lung & H\&E & Lung cancer image classification & Patch based FCN + context-aware block selection and feature aggregation strategy & Private (939 WSIs), TCGA (500 WSIs) \\
\cite{Campanella2019clinical} (\checkmark) & Multi-Cancers & H\&E & Multiple cancer diagnosis in WSIs & CNN (ResNet) + RNNs & Prostate (24,859 slides), skin (9,962 slides), breast cancer metastasis (9,894 slides) \\
\cite{Dov2019deep} & Thyroid & --- & Thyroid malignancy prediction & CNN + ordinal regression for prediction of thyroid malignancy score & Private set (cytopathology 908 WSIs)\\
\cite{Xu2019} (\checkmark) & Multi-Cancers & H\&E & Segmentation of breast cancer metastasis and colon glands & FCN trained on instance-level labels, which are obtained from image-level annotations & Camelyon16 (400 WSIs), Colorectal adenoma private dataset (177 WSIs) \\
\cite{Huang2019celnet} & Breast & H\&E & Localization of cancerous evidence in histopathology images & CNN + multi-branch attention modules and deep supervision mechanism & PCam (327,680 patches extracted from Camelyon16) and Camelyon16 (400 WSIs)\\ \hline \noalign{\vskip 1mm}
\multicolumn{6}{l}{Other approaches} \\ \hline \noalign{\vskip 1mm}
\cite{Campanella2018terabyte} & Prostate & H\&E & Prostate cancer detection & CNN trained under MIL formulation with top-1 ranked instance aggregation approach & Prostate biopsies (12,160 slides) \\
\cite{akbar2018cluster} (\checkmark) & Breast & H\&E & Detection of breast cancer metastasis & Clustering (VAE + K-means) based MIL framework & Camelyon16 (400 WSIs)\\
\cite{tellez2019neural} (\checkmark) & Multi-Cancers & H\&E & Compression of gigapixel histopathology WSIs & Unsupervised feature encoding method (VAE, Bi-GAN, contrastive training) that maps high-resolution image patches to low-dimensional embedding vectors & Camelyon16 (400 WSIs), TUPAC16 (492 WSIs), Rectum (74 WSIs) \\
\cite{Qu2019weakly} (\checkmark) & Multi-Cancers & H\&E & Nuclei segmentation & Modified UNet trained using coarse level-labels + dense CRF loss for model refinement & MoNuSeg (30 images), lung cancer private set (40 images) \\
\cite{Bokhorst19} & Colon & H\&E & Segmentation of tissue types in colorectal cancer & UNet with modified loss functions to circumvent sparse manual annotations & Colorectal cancer WSIs (private set - 70 images) \\
\cite{Li2019weakly} (\checkmark) & Breast & H\&E & Mitosis detection & FCN trained with concentric loss on weakly annotated centriod label & ICPR12 (50 images), ICPR14 (1,696 images), AMIDA13 (606 images), TUPAC16 (107 images)
\\ \bottomrule [0.75pt]
\end{tabularx}
\end{table*}
%%%%%%%%%%%%%%%%%%%%%%%%%%%%%%%%%%%%%%%%%%%%%%%%%%%%%%%%%%%%%%%%%%%%%%%%%%%%%%%%%%
The idea of weakly supervised learning (WSL) is to exploit  coarse-grained (image-level) annotations to automatically infer fine-grained (pixel/patch-level) information. This paradigm is particularly well suited to the histopathology domain, where the coarse-grained information is often readily available in the form of image-level labels, e.g., cancer or non-cancer, but where pixel-level annotations are more difficult to obtain.   Weakly supervised learning dramatically reduces the annotation burden on a pathologist \citep{Xu2014}, and an overview of these models is provided in Table \ref{Tab:Overview of weakly supervised learning models}.

In this survey, we explore one particular form of WSL, namely \textit{multiple-instance learning} (MIL), which aims to train a model using a set of weakly labeled data \citep{Dietterich1997solving, Quellec2017multiple}. In MIL, a training set consists of bags, labeled as positive or negative; and each bag includes many instances, whose label is to be predicted or unknown. For instance, each histology image with cancer/non-cancer label forms a \textit{`bag'} and each pixel/patch extracted from the corresponding image is referred to as an \textit{`instance'} (e.g., pixels containing cancerous cells). Here, the main goal is to train a classifier to predict both bag-level and instance-level labels, while only bag-level labels are  given in the training set. We further categorize MIL approaches into three categories similar to \cite{Cheplygina2019a}: i) \textit{global detection} - identifying a target pattern in a histology image (i.e., at bag level) such as the presence or absence of cancer; ii) \textit{local detection} - identifying a target pattern in an image patch or a pixel (i.e., at instance level) such as highlighting the cancerous tissues or cells; iii) \textit{global and local detection} - detecting whether an image has cancer and also identifying the location where it occurs within an image. These categories are illustrated in Fig. \ref{Fig:An overview of weakly supervised learning models.}. There is also a significant interest in histopathology to include various kinds of weak annotations such as image-level tags \citep{Campanella2019clinical}, points \citep{Qu2019weakly}, bounding boxes \citep{Yang2018boxnet}, polygons \citep{Wang2019} and percentage of the cancerous region within each image \citep{Jia2017}, to obtain clinically satisfactory performance with minimal annotation effort. For an in-depth review of MIL approaches in medical image analysis, refer to \cite{Quellec2017multiple, Cheplygina2019a, Rony2019deep, Kandemir2015}. 

Due to the variable nature of histopathology image appearances, the standard instance-level aggregation methods, such as voting or pooling, do not guarantee accurate image-level predictions, due to misclassifications of instance-level labels \citep{Campanella2019clinical, Rony2019deep}. Hence, several papers on global detection based MIL method rely on alternative instance-level aggregation strategies to obtain reliable bag-level predictions suitable for a given histology task. For instance, \cite{Hou2015efficient} integrated an expectation-maximization based MIL method with a CNN to output patch-level predictions. These instances are later aggregated by training a logistic regression model to classify glioma subtypes in WSIs. \cite{Dov2019deep} proposed an alternative approach based on ordinal regression framework for aggregating instances containing follicular (thyroid) cells to simultaneously predict both thyroid malignancy and TBS score in whole-slide cytopathology images. Recently, a remarkable work in \cite{Campanella2019clinical} adopted an RNN model to integrate semantically rich feature representations across patch-level instances to obtain a final slide-level diagnosis. In their method, the author's managed to obtain an AUC greater than 0.98 in detecting four types of cancers on an extensive multi-centre dataset of 44,732 WSIs, without expensive pixel-wise manual annotations.

The local detection based MIL approaches are based on an image-centric paradigm, where image-to-image prediction is performed using an FCN model - by computing features for all instances (pixels) together. These approaches are generally applied to image segmentation task for precisely delineating cancerous region in histology images. In the local detection approach, the bag labels are propagated to all instances to train a classifier in a supervised manner. However, sometimes even the best bag-level classifier seems to underperform on instance-level predictions due to lack of supervision \citep{Cheplygina2019a}. To tackle this issue, additional weak constraints have been incorporated into FCN models to improve segmentation accuracy. For example, \cite{Jia2017} included an area constraint in the MIL formulation by calculating the rough estimate of the relative size of the cancerous region. However, calculating such area constraints is tedious and can only be performed by an expert pathologist. Consequently, \cite{Xu2019camel} proposed an alternative MIL framework to generate instance-level labels from image-level annotations. These predicted instance-level labels are later assigned to their corresponding image pixels to train an FCN in an end-to-end manner, while achieving comparable performance with supervised counterparts. Finally, in some cases, both a large number of bag labels and a partial set of instance labels are also adopted in FCN based reiterative learning framework \citep{Liang2018weakly}, to further optimize final instance-level predictions.   

Arguably, the most popular and clinically relevant MIL approach in histopathology is the global and local detection paradigm. In this approach, rather than just diagnosing cancer at whole-slide level, we can simultaneously localize the discriminative areas (instances) containing cancerous tissues or cells. In this context, the methods utilize either the bag-level label \citep{Wang2019rmdl} or both bag-level and some coarse level instance annotations \citep{Wang2019} to infer a global level decision. Note that the instance-level predictions are not usually validated due to lack of costly annotations, and are generally visualized as either a heatmap \citep{Wang2019rmdl,Wang2019} or a saliency map \citep{Huang2019celnet} to highlight the diagnostically significant locations in WSIs. The main essence of this approach is to capture the instance-wise dependencies and their impact on the final image-level decision score. 

There is a some disagreement among MIL methods regarding the accuracy of instance-level predictions, when trained with only bag-level labels \citep{Cheplygina2019a, Kandemir2015}. The critical and often overlooked issue among MIL methods is that even the best bag-level classifier may not be an optimal instance-level classifier for instance predictions and vice versa \citep{Cheplygina2019a}. Such problems have naturally led to new solutions that integrate the visual attention models with MIL techniques to enhance the interpretability of final model predictions \citep{Ilse18,Huang2019celnet}. For instance, \cite{Huang2019celnet} proposed a CNN model combining multi-branch attention modules and a deep supervision mechanism \citep{Xie2015holistically}, which aims to localize the discriminative evidence for the class-of-interest from a set of weakly labeled training data. Such attention-based models can precisely pinpoint the location of cancer evidence in WSI, as well as achieving a competitive slide-level accuracy, thereby enhancing the interpretability of current DL models in histopathology applications. 

Not all methods identified as weakly supervised in the literature necessarily fall under the MIL category. For instance, the methods in \cite{Qu2019weakly,Bokhorst19, Li2019weakly} use the term “\textit{weakly supervised}" to indicate that the model training has been performed on sparse set of annotations such as points inside the region of interest \citep{Li2019weakly, Qu2019weakly}, bounding box \citep{Yang2018boxnet} and also some partial pixel-level annotations of cancerous region \citep{Bokhorst19}. These approaches alleviate the need for expensive annotations by proposing newer variants of loss functions \citep{Li2019weakly}, feature encoding strategies \citep{tellez2019neural, akbar2018cluster}, loss balancing mechanisms \citep{Bokhorst19}, and methods to derive coarse labels from weak annotations \citep{Qu2019weakly} in order to eventually train fully-supervised models in a weakly supervised way. 

\subsection{Unsupervised learning}
\label{ssec:Unsupervised methods}

\begin{figure*}\centering
\centerline{\includegraphics[width=0.75\linewidth]{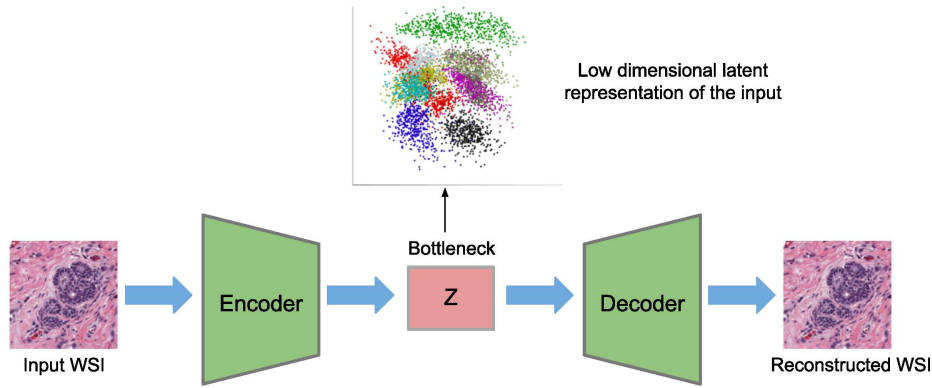}}
\caption{An overview of unsupervised learning models.}
\label{Fig:An overview of unsupervised learning models.}
\end{figure*}
The goal of unsupervised learning is to learn something useful about the underlying data structure without the use of labels. The term “unsupervised" is sometimes used loosely among the digital pathology community for approaches that are not fully unsupervised. For instance, stain transfer without pairing, or domain adaptation via feature distribution matching are considered as unsupervised, even though the domains can be considered as labels for two separate datasets \citep{gadermayr2019generative, de2019stain, ganin2016domain}. In this survey, we examine fully unsupervised methods, where the raw data comes in the form of images without any identifiers (e.g., domain, cancerous vs. non-cancerous, tissue etc.). These approaches are rare, since the field of unsupervised learning among the machine learning community is also still in its infancy. However, it is clear why one should be interested in such approaches as the scarcity of labeled data due to regulatory concerns and labor costs (i.e., expert annotations) is a major bottleneck in achieving clinically satisfactory performance in medical imaging \citep{lee2017medical}. 

In unsupervised learning, the learning task is ambiguous, since it is possible to map the inputs into infinitely many subsets, provided there are no restrictions. Most unsupervised approaches aim to maximize the probability distribution of the data, subject to some constraints, in order to limit the solution space and to achieve a desired grouping/clustering for the target task. A common technique is to transform the data into a lower-dimensional subspace, followed by aggregation of feature representations into mutually exclusive or hierarchical clusters, which is illustrated in Fig. \ref{Fig:An overview of unsupervised learning models.}. Autoencoders are typically utilized for the dimensionality reduction step. Recent  advances in modeling the stochasticity \citep{kingma2013auto}, and more robustly disentangling visual features \citep{Higgins2017betaVAELB, chen2018isolating} have made autoencoders more attractive for feature modeling and dimensionality reduction. In early work, sparse autoencoders were utilized for unsupervised nuclei detection \citep{xu2015stacked}. Later, detection performance was improved by modifying the receptive field of the convolutional filters to accommodate small nuclei \citep{hou2019sparse}. For more complex tasks, such as tissue and cell classification, Generative Adversarial Networks (GANs) have also been employed. Specifically, InfoGANs \citep{chen2016infogan} have been used for extracting features, which  maximize the mutual information between the generated images and a predefined subset of latent (noise) codes, which are then used for tasks such as cell-level classification, nuclei segmentation, and cell counting \citep{hu2018unsupervised}. 

Finally, we examine unsupervised transfer learning approaches, where instead of directly applying learned features on a target task, learned mapping functions are used as an initialization for target tasks, possibly with very few labeled training images. Using a loss term that is similar to the reconstruction objective of autoencoders, \citep{chang2017unsupervised} trains a convolutional network using unlabeled images pertaining to a specific modality (e.g., brain MRI or kidney histology images), to learn filter banks at different scales. The resulting filters are shift-invariant, scale-specific, and can uncover intricate patterns in various tasks, such as tumour classification of glioblastoma multiforme or kidney renal clear cell carcinoma. In machine learning, this form of unsupervised learning is called \textit{``self-supervised"} learning. Since self-supervised techniques can deal with larger images in general, they offer a promising alternative to clustering approaches in histopathology, which usually require context and a larger field of view. Context-based self-supervised methods which predict spatial ordering \citep{noroozi2016unsupervised} or image rotations \citep{gidaris2018unsupervised}, and generative methods such as mapping grayscale images into their RGB counterparts have been successfully used for initializing networks for faster convergence and learning target tasks with fewer labels. However, in histopathology, the rules governing the spatial location of cell structures, or the color or staining of a histology image are different to those for natural scene images. While, this makes the task of unsupervised learning more difficult for histopathology images, it also presents an opportunity for researchers to develop novel techniques that may be applicable to medical images. 

Unsupervised learning methods are desirable as they allow models to be trained with little or no labeled data. Furthermore, as these methods are constructed to disentangle relationships between samples in the dataset for grouping (or clustering), a successful unsupervised learning method can also improve the interpretability of a model, by examining how the model groups items into separate categories. While fully unsupervised methods for arbitrary tasks are still uncommon, techniques used for auxiliary tasks (e.g., pre-training) such as self-supervision \citep{tellez2019neural} can reduce the annotation burden on the expert, thereby significantly expediting the research.
%%%%%%%%%%%%%%%%%%%%%%%%%%%%%%%%%%%%%%%%%%%%%%%%%%%%%%% Table %%%%%%%%%%%%%%%%%%%%%%%%%%%%%%%%%%%%%%%%%%%%%%%%%%%%%%%%%%%%%%%%
\begin{table*}
\scriptsize
\centering
\caption{Overview of unsupervised learning models. Note: (\checkmark) indicates the code is publicly available and the link is provided in their respective paper.}
\label{Tab:Overview of unsupervised learning models}
\begin{tabularx}{\textwidth}{lllXXX}
\toprule[0.75pt]
Reference & \multicolumn{1}{l}{Cancer types} & \multicolumn{1}{l}{Staining} & \multicolumn{1}{l}{Application} & \multicolumn{1}{l}{Method} & \multicolumn{1}{l}{Dataset} \\ \midrule 
\cite{xu2015stacked} & Breast & H\&E & Nuclei segmentation & Stacked sparse autoencoders & 537 H\&E images from Case Western Reserve University \\
\cite{hu2018unsupervised} (\checkmark) & Bone marrow & H\&E & Tissue and cell classification & InfoGAN & 3 separate datasets: public data with 11 patches of size $1200\times1200$, private datasets with WSIs of 24 patients + 84 images \\
\cite{Bulten18} & Prostate & H\&E, IHC & Classification of prostate into tumour vs. non-tumour & Convolutional adversarial autoencoders & 94 registered WSIs from Radboud University Medical Center\\
\cite{Sari2019UnsupervisedFE} & Colon & H\&E, IHC & Subtyping of intrahepatic cholangiocarcinoma (ICC) & Restricted Boltzmann Machines + Clustering & 3236 images, private dataset \\
\cite{Quiros2019PathologyGL} (\checkmark) & Breast & H\&E & High resolution image generation + feature extraction & BigGAN + Relativistic GAN & 248 + 328 patients from private dataset \\ 
\cite{hou2019sparse} & Breast & H\&E & Nuclei detection, segmentation and representation learning & Sparse autoencoder & 0.5 million images of nuclei from TCGA \\
\cite{gadermayr2019generative} & Kidney & Stain agnostic & Segmentation of object-of-interest in WSIs & CycleGAN + UNet segmentation & 23 PAS, 6 AFOG, 6 Col3 and 6 CD31 WSIs\\
\cite{de2019stain} & Kidney & Stain agnostic & Tissue segmentation & CycleGAN + UNet segmentation & Private set containing 40 + 24 biopsy images \\
\cite{pmlr-v102-gadermayr19a} & Kidney & PAS, H\&E & Segmentation of the glomeruli & CycleGAN & 23 WSIs, private dataset
\\ \bottomrule [0.75pt]
\end{tabularx}
\end{table*}

%%%%%%%%%%%%%%%%%%%%%%%%%%%%%%%%%%%%%%%%%%%%%%%%%%%%%%%%%%%%%%%%%%%%%%%%%%%%%%%%%

\subsection{Transfer learning}
\label{ssec:Transfer learning}

\begin{table*}
\scriptsize
\centering
\caption{Overview of transfer learning models. Note: (\checkmark) indicates the code is publicly available and the link is provided in their respective paper.}
\label{Tab:Overview of transfer learning models}
\begin{tabularx}{\textwidth}{lllXXX}
\toprule[0.75pt]
Reference & \multicolumn{1}{l}{Cancer types} & \multicolumn{1}{l}{Staining} & \multicolumn{1}{l}{Application} & \multicolumn{1}{l}{Method} & \multicolumn{1}{l}{Dataset} \\ \midrule 
\cite{Wang2016b} & Breast & H\&E & Detection of breast cancer metastasis & Pre-trained GoogleNet model & Camelyon16 (400 WSIs) \\
\cite{Liu2017} & Breast & H\&E & Detection of breast cancer metastasis & Pre-trained Inception-V3 model & Camelyon16 (400 WSIs) \\
\cite{Han2017} & Breast & H\&E & Breast cancer multi-classification & CNN integrated with feature space distance constraints for identifying feature space similarities & BreaKHis (7,909 images)\\
\cite{Lee2018} & Breast & H\&E & Detection and pN-stage classification of breast cancer metastasis & Patch based CNN for metastasis detection + Random forest classifier for lymph node classification & Camelyon17 (1,000 WSIs) \\
\cite{Chennamsetty2018} & Breast & H\&E & Breast cancer classification & Ensemble of three pre-trained CNNs + aggregation using majority voting & BACH 2018 challenge (400 WSIs) \\
\cite{Kwok2018} & Breast & H\&E & Breast cancer classification & Inception-Resnet-V2 based patch classifier & BACH 2018 challenge (400 WSIs) \\
\cite{Bychkov2018} & Colon & H\&E & Outcome prediction of colorectal cancer & A 3-layer LSTM + VGG-16 pre-trained features to predict colorectal cancer outcome & Private set (420 cases) \\
\cite{Arvaniti2018} (\checkmark) & Prostate & H\&E & Predicting Gleason score & Pre-trained MobileNet architecture & Private set (886 cases) \\
\cite{Coudray2018} (\checkmark) & Lung & H\&E & Genomics prediction from pathology images & Patch based Inception-V3 model & TCGA (1,634 WSIs) and validated on independent private set containing frozen sections (98 slides), FFPE sections (140 slides) and lung biopsies (102 slides) \\
\cite{Kather2019} (\checkmark) & Colon & H\&E & Survival prediction of colorectal cancer & Pre-trained VGG-19 based patch classifier & TCGA (862 WSIs) and two other public datasets (25 + 86 WSIs)\\
\cite{Noorbakhsh2019} (\checkmark) & Multi-Cancers & H\&E & Pan-cancer classification & Pre-trained Inception-V3 model & TCGA (27,815 WSIs) \\
\cite{Tabibu2019} (\checkmark) & Kidney & H\&E & Classification of Renal Cell Carcinoma subtypes and survival prediction & Pre-trained ResNet based patch classifier & TCGA (2,093 WSIs) \\
\cite{akbar2019automated} & Breast & H\&E & tumour cellularity (TC) scoring & Two separate InceptionNets: one for classification (healthy vs. cancerous tissue) and the other outputs regression scores for TC & BreastPathQ (96 WSIs) \\
\cite{Valkonen2019} (\checkmark) & Breast & ER, PR, Ki-67 & Cell detection & Fine-tuning partially pre-trained CNN network & DigitalPanCK (152 - invasive breast cancer images) \\
\cite{strom2020artificial} & Prostate & H\&E & Grading of prostate cancer & Ensembles of two pre-trained Inception-V3 models & Private set (8730 WSI's)
\\ \bottomrule [0.75pt]
\end{tabularx}
\end{table*}
The most popular and widely adopted technique in digital pathology is the use of \textit{transfer learning} approach. In transfer learning, the goal is to extract knowledge from one domain (i.e., source) and apply it to another domain (i.e., target) by relaxing the assumption that the train and test set must be independent and identically distributed. In histopathology, transfer learning is typically done using ImageNet pretrained models such as \textit{VGGNet} \citep{Simonyan2014}, \textit{InceptionNet} \citep{Szegedy2015,Szegedy2016}, \textit{ResNet} \citep{He2016}, \textit{MobileNet} \citep{Howard2017}, \textit{DenseNet} \citep{Huang2017}, and various other variants of these models. These pre-trained models have been widely applied to various cancer grading and prognosis tasks (Refer, Table \ref{Tab:Overview of transfer learning models} and Section \ref{sec:Survival models for disease prognosis} for more details). A critical analysis of best-performing methods on various Grand Challenges is discussed thoroughly in Section \ref{sec:Critical analysis of architectures}.

In digital pathology, different types of staining are used depending on the application. Immunohistochemistry (IHC)  allows specific molecular targets to be visualized (e.g., Ki-67 to estimate tumour cell proliferation rate \citep{Valkonen2019}, and cytokeratin to detect micrometastases \citep{Bejnordi2017jama}, whilst H\&E is a widely-used general-purpose stain. The appearance of images varies widely depending on the stain used and also on the degree of staining, and this poses a unique challenge, as CNN's are highly sensitive to the data they were trained on \citep{ciompi2017importance}. In the following sub-sections, we review two approaches used to overcome this problem.

\subsubsection{Domain adaptation}
\label{sssec:Domain adaptation}

\begin{table*}
\scriptsize
\centering
\caption{Overview of domain adaptation and stain normalization models. Note: (\checkmark) indicates the code is publicly available and the link is provided in their respective paper.}
\label{Tab:Overview of domain adaptation methods}
\begin{tabularx}{\textwidth}{lllXXX}
\toprule[0.75pt]
Reference & \multicolumn{1}{l}{Cancer types} & \multicolumn{1}{l}{Staining} & \multicolumn{1}{l}{Application} & \multicolumn{1}{l}{Method} & \multicolumn{1}{l}{Dataset} \\ \midrule 
\multicolumn{6}{l}{Domain adaptation} \\ \hline \noalign{\vskip 1mm}
\cite{lafarge2017domain} & Breast & H\&E & Mitosis detection & Gradient reversal with CNNs & TUPAC16 (73 WSIs) \\
\cite{ren2018adversarial} & Prostate & H\&E & Feature matching of image patches & Siamese networks & TCGA + private dataset \\
\cite{brieu2019domain} & Multi-Cancers & Multi-stain & Semi-automatic nuclei labeling using stain transfer & CycleGAN & TCGA (75 bladder cancer + 29 lung cancer + 142 tissue samples of FOVs images)  + 30 FOVs of breast cancer (private set) \\
\cite{gadermayr2019generative} & Kidney & Stain agnostic & Segmentation of object-of-interest in WSIs & CycleGAN + UNet segmentation & 23 PAS, 6 AFOG, 6 Col3 and 6 CD31 WSIs \\
\cite{DBLP:journals/corr/abs-1906-11118} & Lung & PD-L1 + Cytokeratin & Segmentation & CycleGAN + SegNet segmentation model & 56 Cytokeratin + 69 PD-L1 WSIs (private set)\\
\cite{ciga2019multi} & Breast &  H\&E & Classification & Multi-layer gradient reversal & BACH\\ 
\hline \noalign{\vskip 1mm}
\multicolumn{6}{l}{Stain variability} \\ \hline \noalign{\vskip 1mm}
\cite{janowczyk2017stain} (\checkmark) & Multi-Cancers & H\&E & Stain transfer for H\&E staining & Sparse autoencoders & 5 breast biopsy slides + 7 gastro-intestinal biopsies\\
\cite{cho2017neural} & Breast & H\&E & Stain transfer & DCGAN conditioned on a target image & CAMELYON16 \\
\cite{bentaieb2017adversarial} & Multi-Cancers & H\&E & Stain transfer & GAN + regularization based on auxiliary task performance & ICPR2014 + GLAS challenge + 135 WSIs (private set) \\
\cite{zanjani2018histopathology} (\checkmark) & Lymph nodes & H\&E & Stain transfer for H\&E staining & Multiple studies with Gaussian mixture models, variational autoencoders, and InfoGAN & 625 images from 125 WSIs of lymph nodes from 3 patients \\
\cite{de2019stain} & Kidney & Stain agnostic & Segmentation & CycleGAN + UNet segmentation & 40 + 24 biopsy images (private) \\
\cite{shaban2019staingan} (\checkmark) & Breast & H\&E & Stain transfer & CycleGAN & ICPR2014 \\
\cite{rivenson2019phasestain} & Multi-Cancers & H\&E, Jones, Masson’s trichrome & Digital staining of multiple tissues & Custom GAN & N/A \\
\cite{lahiani2019virtualization} & Liver & FAP-CK from Ki67-CD8 & Virtual stain transformation between different types of staining & CycleGAN + Instance normalization & 10 Ki67-CD8 + 10 FAP-CK stained colorectal carcinoma WSIs 
\\ \bottomrule [0.75pt]
\end{tabularx}
\end{table*}
 
Domain adaptation is a sub-field of \textit{transfer learning}, where a task is learned from one or more source domains with labeled data, and the aim is to achieve similar performance on the same task on a target domain with little or no labeled data \citep{wang2018deep}. Domain-adversarial networks are designed to learn features that are discriminative for the main prediction task whilst being insensitive to domain shift \citep{ganin2016domain, lafarge2017domain, ren2018adversarial} and this approach has been applied to digital pathology. \cite{ciga2019multi} achieved state-of-the-art performance on the BACH (BreAst Cancer Histopathology) challenge task using a multi-level domain-adversarial network. \cite{ren2018adversarial} performed unsupervised training based on siamese networks on prostate WSIs, positing that given a WSI, different patches should be given the same Gleason score, thereby extracting common features present in different parts of the WSI. This auxiliary task also helped increase the adversarial domain adaptation performance on another target dataset.
 
Fake (artificially generated) images are also used in domain adaptation. \cite{brieu2019domain} utilized semi-automatic labeling of the nuclei with one type of staining (IHC) to alleviate the costlier annotation of another staining method (H\&E), where fake H\&E images are generated from IHC images to increase the dataset size. Similarly, \cite{gadermayr2019generative} used artificial data generation with GANs for semantic segmentation in kidney histology images with multiple stains. Each work uses adversarial models (i.e., generators and discriminators) for image-to-image translation utilizing cycle consistency loss for unpaired training. The translation is performed to obtain an intermediate, stain-agnostic representation \citep{lahiani2019virtualization}, which is then fed to a network trained on this representation to perform segmentation.

\subsubsection{Stain normalization}
\label{sssec:Stain normalization}

Stain normalization, augmentation and stain transfer are popular image preprocessing techniques to improve generalization of a task by modifying the staining properties of a given image to match another image visually. In contrast to the methods described in Section \ref{sssec:Domain adaptation}, which modify the \textit{features} extracted from different image distributions so that they are indistinguishable from each other; stain normalization directly modifies the input images to obtain features that are invariant to staining variability.

One may combat staining variation by augmenting the training data by varying each pixel value per channel within a predefined range on transformed color spaces, such as HSV (hue, saturation and value) or HED (Hematoxylin, Eosin, and Diaminobenzidine) \citep{Liu2017, Li2018, Tellez2018}. Earlier machine learning (ML) methods \citep{macenko2009method, vahadane2016structure} assume that staining attenuates light (optical density) uniformly and decompose each optical density image into concentration (appearance) and color (stain) matrices. The uniformity assumption is relaxed in more recent ML methods, where the type of chemical staining and morphological properties of an image are considered in generating stain matrices \citep{khan2014nonlinear, bejnordi2015stain}. Neural networks, such as sparse autoencoders for template matching \citep{janowczyk2017stain}, and GANs are also used for stain transfer and normalization \citep{zanjani2018histopathology, de2019stain, bentaieb2017adversarial, cho2017neural}. Cycle consistency loss objective \citep{zhu2017unpaired} has been utilized for improved stain transfer with structure preservation, as well as for training systems without annotating the pairing between the source (to be stained) and the target (used as a reference for staining new images) \citep{shaban2019staingan, de2019stain, cho2017neural}. Auxiliary tasks, such as maintaining high prediction accuracy on classification or segmentation, have led to consistent stain transfer accounting for the type or shape of the tissue present \citep{odena2017conditional, bentaieb2017adversarial}. Recently, the same techniques have also been used for virtually staining quantitative phase images of label-free tissue sections \citep{rivenson2019phasestain}.

Although stain transfer methods produce aesthetically pleasing results, their use cases are still not entirely clear. For instance, \cite{shaban2019staingan} report considerable gains compared to a baseline (no augmentations or traditional methods such as \cite{macenko2009method, reinhard2001color, vahadane2016structure} in the CAMELYON16 challenge; however, the winning entry (by a margin of 21\% with respect to \cite{shaban2019staingan} in AUC for a binary classification task on WSIs) utilizes a traditional machine learning based normalization technique that aligns chromatic and density distributions of source and the target \citep{bejnordi2015stain}. A thorough study comparing numerous approaches found that it is always advisable to apply various forms of color augmentation in HSV or HED space, and additional slight performance gains are still achievable with a network-based augmentation strategy \citep{tellez2019quantifying}. Similarly, \cite{DBLP:journals/corr/abs-1909-11575} examined the effect of domain shift on histopathological data and automated medical imaging systems, and found that augmentation and normalization strategies drastically alter the performance of these systems. Differentiating factors such as scale-spaces, resolution, image quality, scanner imperfections are also likely to affect the performance of a model, in addition to staining, which is less explored in the community.

\section{Survival models for disease prognosis}
\label{sec:Survival models for disease prognosis}

This section concentrates on methods of training survival models that can either generate a probability of an event in a certain predefined period of time, or can predict time to an event using regression from a WSI. In the context of cancer, the term prognosis refers to the likely outcome for a patient on standard treatment, and the term prediction refers to how the patient responds to a particular treatment. Since the difference between these two terms is not relevant when carrying out survival analysis, we will use the term prediction to cover both prediction and prognosis.  The outcome metrics used to train a prediction model will depend on the disease. For example, in patients with very aggressive disease such as glioblastoma, the survival time in months may be used as an endpoint, whereas for breast cancer, with an average survival rate at 10 years of around 80\%, the time to recurrence of the disease after surgery is a more relevant metric and at least 5 years follow-up is required. Following up patients prospectively is a time consuming and expensive process and for this reason several studies use existing clinically validated risk models, or genomics assays as a proxy for long term outcomes; for example, the use of PAM50 scores \citep{VETA2019111, couture2018image} in breast cancer and Gleason grades in prostate cancer \citep{Nagpal2019}. The survival data or risk scores may be dichotomized e.g., survival at specific time points or risk score above or below a set cutoff; this allows the survival model to be treated as a classification problem, but information is lost and new models have to be trained if the cutoff value is changed, e.g., two different models are needed to predict 5-year and 10-year survival times. Time to event models are more complicated since nothing is known about what happens to a patient after they are lost to followup; this is known as right censoring.  A proportional hazards model is commonly used to model an individual’s survival and can be implemented using a neural network \citep{katzman2018deepsurv} and several groups have used this approach in digital pathology \citep{zhu2017wsisa,mobadersany2018predicting, tang2019capsurv}.  

The data used to train survival models is weakly labeled, with only one outcome label per patient. This poses a computational challenge as a WSI is so large that it has to be broken down into 100’s or even 1000’s of smaller patches for processing; since tumours may be very heterogeneous, only a subset of these patches may be salient for the prediction task.  Although \cite{Campanella2019clinical} recently demonstrated that a relatively simple MIL approach could produce accurate diagnostic results when more than 10,000 slides were available for training, datasets for survival analysis usually have fewer than 1000 slides available which makes the task much more difficult. Three main approaches are used to overcome the shortage of labeled data. The first is to use the image features that expert pathologists have already identified as being associated with survival. Examples include assessing tumour proliferation \citep{VETA2019111} in breast cancer, quantifying the stroma/tumour ratio in colorectal cancer \citep{geessink2019computer} and predicting the Gleason grade in prostate cancer \citep{Nagpal2019}. The role of deep learning in these cases is to provide an automatic and reproducible method for extracting these features; a vital advantage of this approach over end-to-end models is that results are more interpretable since each component can be assessed individually. In the second approach, image features are extracted from image patches using a pre-trained CNN, then feature selection or dimensionality reduction is carried out and finally a survival model is trained on the resulting feature vector. Examples include survival prediction in mesothelioma \citep{courtiol2019deep} and colorectal cancer \citep{Kather2019, Bychkov2018}, and risk of recurrence in breast cancer \citep{couture2018image}. In the third approach, unsupervised methods are used to learn a latent representation of the data which is then used to train the survival model. For example, \cite{zhu2017wsisa} apply K-means to small patches to identify 50 clusters or phenotypes for glioma and non-small-cell lung cancer, and \cite{muhammad2019} use an autoencoder with an additional clustering constraint to predict survival in intrahepatic cholangiocarcinoma. 

There are many possible ways of aggregating the predictions for individual patches to give a single prediction for a patient. The simplest approach is to take the mean prediction across all patches \citep{tang2019capsurv}, but this will not work if the salient patches only represent  a small fraction of the WSI; for this reason other schemes, such as taking the average of the two highest ranking patches \citep{mobadersany2018predicting}, may be more appropriate. Some methods generate a low dimensional feature vector that captures the distribution of scores across the patches. For example, \cite{Nagpal2019} use the distribution of Gleason scores across all patches as an input feature vector to a KNN to generate a patient score, and \cite{couture2018image} aggregate patch probabilities into a quantile function which is then used by an SVM to generate a patient level class. Methods that assign patches to discrete classes or clusters can simply use the majority class to label the WSI \citep{muhammad2019} or adopt a RNN to generate a single prediction from a sequence of patches \citep{Bychkov2018}. 

End-to-end methods that learn features directly from the image data and allow probabilities to be associated with individual patches can be used to uncover new information how morphology is related to outcome. For example, \cite{courtiol2019deep} were able to show that regions associated with stroma, inflammation, cellular diversity, and vacuolization were important in predicting survival in mesothelioma patients, and \cite{mobadersany2018predicting} showed that microvascular proliferation and increased  cellularity is associated with poorer outcome in glioma patients. Prediction heatmaps may also allow researchers to uncover patterns of tumour  heterogeneity and could be used to guide  tissue extraction for genomics and proteomics assays. Deep learning survival models are, therefore, of great interest to  cancer researchers as well as to pathologists and oncologists. 

%%%%%%%%%%%%%%%%%%%%%%%% Table %%%%%%%%%%%%%%%%%%%%%%%%%%%%%%%%%%%%%%%
\begin{table*}
\scriptsize
\centering
\caption{Overview of survival models for disease prognosis. Note: (\checkmark) indicates the code is publicly available and the link is provided in their respective paper.}
\label{Tab:Overview of survival models for disease prognosis.}
\begin{tabulary}{\textwidth}{llLLL}
\toprule[0.75pt]
Reference & \multicolumn{1}{l}{Cancer types} &  \multicolumn{1}{l}{Application} & \multicolumn{1}{l}{Method} &
\multicolumn{1}{l}{Dataset} \\ \midrule 

\cite{zhu2017wsisa} & Multi-Cancers & Loss function based on survival time & Raw pixel values of downsampled patches used as feature vectors; 10 clusters identified using K-means clustering. Deep survival models are trained for each cluster separately. Significant clusters are identified and corresponding scores are fed into final WSI classifier & TCIA-NLST, TCGA-LUSC, TCGA-GBM \\ 

\cite{Bychkov2018} & Colorectal & 5 year disease specific survival & Extracted features using pre-trained VGG-16. Used RNN to generate WSI prediction from tiles & Private set - TMAs from 420 patients \\

\cite{couture2018image} (\checkmark) & Breast & Prediction of tumour grade, ER status, PAM50 intrinsic subtype, histologic subtype and risk of recurrence score & Pre-trained VGG-16 model. Aggregate features over $800 \times 800$ regions to predict class for each patch, then frequency distribution of classes input to SVM to combine regions to predict TMA class & TMA cores (Private-1203 cases) \\

\cite{mobadersany2018predicting} (\checkmark) & Brain & Time to event modelling & CNN integrated with a Cox proportional hazards model to predict patient outcomes using histology and genomic biomarkers. Calculate median risk for each ROI, then average 2 highest risk regions & TCGA-LGG, TCGA-GBM (1,061 WSIs)\\

\cite{courtiol2019deep} & Mesothelioma  & Loss function based on survival time & Pre-trained ResNet50 extracts features from 10000 tiles. 1-D convolutional layer generates score for each tile. 10 highest and lowest scores fed into MLP classifier for WSI prediction & MESOPATH/MESOBANK (private set-2,981 WSIs), TCGA validation set (56 WSIs)\\

\cite{geessink2019computer} & Colorectal & Dichotomized tumour/stromal ratios & CNN based patch classifier trained to identify tissue components. Calculate tumour-stroma ratio for manually defined hot-spots & Private set-129 WSIs \\

\cite{Kather2019} (\checkmark) & Colorectal & Dichotomized stromal score & VGG-19 based patch classifier trained to identify tissue component. Calculate HR for each tissue component using mean activation. Combine components with $HR>1$ to give a “deep stromal score" & NCT-CRC-HE-100k; TCGA-READ, TCGA-COAD \\

\cite{muhammad2019} & Liver ICC & HRs of clusters compared & Unsupervised method to cluster tiles using autoencoder. WSI assigned to cluster corresponding to majority of tiles & Private set - 246 ICC H\&E WSIs \\

\cite{Nagpal2019} & Prostate & Gleason scoring & Trained Inception-V3 network to predict Gleason score on labeled patches. Then calculate \% patches with each grade on the WSI and use result as a low dimensional feature vector input to k-NN classifier & TCGA-PRAD and private dataset \\

\cite{qaiser2019digital} & Lymphoma & Generate 4 DPC categories & Multi-task CNN model for simultaneous cell detection and classification, followed by digital proximity signature (DPS) estimation & Private set-32 IHC WSIs \\

\cite{tang2019capsurv} & Multi-Cancers & Dichototomized survival time ($<= 1$ year and $> 1$ year) & A capsule network is trained using a loss function that combines a reconstruction loss, margin loss ans Cox loss. The mean of all patch-level survival predictions is calculated to achieve a final patient-level survival prediction. & TCGA-GBM and TCGA-LUSC \\

\cite{VETA2019111} & Breast & Predict mitotic score \& PAM50 proliferation score & Multiple methods from challenge teams & TUPAC 2016 \\

\cite{yamamoto2019automated} & Prostate & Predict accuracy of prostate cancer recurrence & Deep autoencoders trained at different magnifications and weighted non-hierarchical clustering, followed by SVM classifier to predict the short-term biochemical recurrence of prostate cancer & Private set - 15,464 WSI's

\\ \bottomrule [0.75pt]
\end{tabulary}
\end{table*}

\section{Discussion and future trends}
\label{sec:Discussion}

%% Datasets %%%%%%%%%%%%%%%%%%%%%%%%%%%%%%
\begin{table*}
\tiny
\centering
\caption{Summary of publicly available databases in computational histopathology.}
\label{Tab:Summary of publicly available datasets}
\begin{tabularx}{\textwidth}{p{2.2cm}p{0.5cm}p{2cm}p{2cm}p{3cm}X}
\toprule[0.75pt]
Dataset / Year & \multicolumn{1}{l}{Cancer types} & \multicolumn{1}{l}{Goal} & \multicolumn{1}{l}{Images / Cases (train+test)} & \multicolumn{1}{l}{Annotation} & \multicolumn{1}{l}{Link} \\ \midrule 
ICPR 2012 \citep{Cirecsan2013} & Breast & Mitosis detection & 50 (35+15) & Pixel-level annotation of mitotic cells & \url{http://ludo17.free.fr/mitos_2012/}\\
AMIDA 2013 \citep{veta2015assessment} & Breast & Mitosis detection & 23 (12+11) & Centroid pixel of mitotic cells & \url{http://amida13.isi.uu.nl/} \\
ICPR 2014 \citep{Cirecsan2013} & Breast & Mitosis detection & 2112 (2016+96) & Centroid pixel of mitotic cells & \url{https://mitos-atypia-14.grand-challenge.org/}\\
GLAS 2015 \citep{Sirinukunwattana2017} & Colon & Gland segmentation & 165 (85+80) & Glandular boundaries & \url{https://warwick.ac.uk/fac/sci/dcs/research/tia/glascontest/} \\
TUPAC 2016 \citep{VETA2019111} & Breast & tumour proliferation based on mitosis counting \& molecular data + two auxiliary tasks & 821 (500+321) + (73/34) & Proliferation scores \& ROI of mitotic cells & \url{http://tupac.tue-image.nl/} \\
HER2 Scoring 2016 \citep{Qaiser2018her} & Breast & HER2 scoring in breast cancer WSIs & 86 (52+28) & HER2 score on whole-slide level & \url{https://warwick.ac.uk/fac/sci/dcs/research/tia/her2contest/} \\
BreakHis 2016 \citep{spanhol2015dataset} & Breast & Breast cancer detection & 82 (7909 patches) & WSL benign vs. malignant annotation & \url{https://web.inf.ufpr.br/vri/databases/breast-cancer-histopathological-database-breakhis/}\\
CRCHisto 2016 \citep{Sirinukunwattana2016} & Colon & Nuclei detection \& classification & 100 & 29,756 nuclei centres + out of which 22,444 with associated class labels & \url{https://warwick.ac.uk/fac/sci/dcs/research/tia/data/crchistolabelednucleihe} \\
CAMELYON16 \citep{Bejnordi2017} & Breast & Breast cancer metastasis detection & 400 (270+130) & Contour of cancer
locations & \url{https://camelyon16.grand-challenge.org/} \\
CAMELYON17 \citep{Bandi2018} & Breast & Breast cancer metastasis detection \& pN-stage prediction & 1000 (500+500) & Contour of cancer
locations + patient level score & \url{https://camelyon17.grand-challenge.org/} \\
MoNuSeg 2018 \citep{Kumar2019} & Multi-Cancers & Nuclei segmentation & 44(30+14) & 22,000+7000 nuclear boundary annotations & \url{https://monuseg.grand-challenge.org/Home/} \\
PCam 2018 \citep{veeling2018rotation} & Breast & Metastasis detection & 3,27,680 patches & Patch-level binary label & \url{https://github.com/basveeling/pcam} \\
TNBC 2018 \citep{Naylor2018} & Breast & Nuclei segmentation & 50 & 4022 pixel-level annotated nuclei & \url{https://github.com/PeterJackNaylor-/DRFNS} \\
BACH 2018 \citep{Aresta2019} & Breast & Breast cancer classification & 500 (400+100) & Image-wise \& pixel-level annotations & \url{https://iciar2018-challenge.grand-challenge.org/Home/} \\
BreastPathQ 2018 \citep{akbar2019automated} & Breast & tumour cellularity & 96 (69+25) WSIs & 3,700 patch-level tumour cellularity score & \url{https://breastpathq.grand-challenge.org/} \\ 
Post-NAT-BRCA \citep{Martel2019tcia} & Breast & tumour cellularity &  96 WSIs & Nuclei, patch and patient level annotations & \url{https://doi.org/10.7937-/TCIA.2019.4YIBTJNO} \\
CoNSeP 2019 \citep{Graham2019} & Colon & Nuclei segmentation and classification & 41 & 24,319 pixel-level annotated nuclei & \url{https://warwick.ac.uk/fac/sci/dcs-/research/tia/data/hovernet/} \\
CRAG 2019 \citep{Graham2019mild} & Colon & Gland segmentation & 213 (173+40) & Gland instance-level ground truth & \url{https://warwick.ac.uk/fac/sci/dcs-/research/tia/data/mildnet/} \\
LYON 2019 \citep{Swiderska2019} & Multi-Cancers & Lymphocyte detection & 83 WSIs & 171,166 lymphocytes in 932 ROIs were annotated & \url{https://lyon19.grand-challenge.org/Home/} \\
NCT-CRC-HE-100k \citep{Kather2019} & Colon & Tissue classification & 1,00,000 patches(86 WSIs)+7,180 patches(25 WSIs) & Patch-label for nine class tissue classification  & \url{https://zenodo.org/record/1214456#.XffRa3VKhhE} \\ 
ACDC-LungHP 2019 \citep{li2018computer} & Lung & Detection and classification of lung cancer subtypes & (150 + 50) WSI's & Contour of cancer locations + image-level cancer subtype scores & \url{https://acdc-lunghp.grand-challenge.org/}\\
Dataset of segmented nuclei 2020 \citep{hou2020dataset} & Multi-Cancers & Nuclei segmentation & 5,060 (WSI's) + 1,356 (image patches) & Patch level labels (1,356 patches) + ~ 5 billion pixel-level nuclei labels & \url{https://wiki.cancerimagingarchive.net/display/DOI/Dataset+of+Segmented+Nuclei+in+Hematoxylin+and+Eosin+Stained+Histopathology+Images} \\
TCGA \citep{TCGA} & Multi-Cancers & Multiple & ---- & ---- & \url{https://portal.gdc.cancer.gov/} \\
TCIA \citep{TCIA} & Multi-Cancers & Multiple & ---- & ---- & \url{https://www.cancerimagingarchive.net/}\\ 
\bottomrule [0.75pt]
\end{tabularx}
\end{table*}

\subsection{Effect of deep learning architectures on task performance}
\label{sec:Critical analysis of architectures}

In most applications, standard architectures (e.g., \textit{VGGNet} \citep{Simonyan2014}, \textit{InceptionNet} \citep{Szegedy2015,Szegedy2016}, \textit{ResNet} \citep{He2016}, \textit{MobileNet} \citep{Howard2017}, \textit{DenseNet} \citep{Huang2017}) can be directly employed, and custom networks should only be used if it is impossible to transform the inputs into a suitable format for the given architecture, or the transformation may cause significant information loss that may affect the task performance. For instance, if the scanner pixel scale does not match with the powers of two for nuclei segmentation, custom neural networks with varying image sizes (e.g., $71\times71$) can be utilized \citep{saha2017advanced}. The most standard architectures are exhaustively tested by many, where their pitfalls, convergence behaviour and weaknesses are well documented in the literature. Unlike the previous, the custom network design choices such as the type of pooling - performed for spatial dimensionality reduction, sizes of the convolutional filters, inclusion of residual connections and/or any other blocks (e.g., Squeeze-and-Excite modules \citep{hu2018squeeze} or inception modules \citep{Szegedy2016}) are left for the researchers to explore and can be critical to the performance of the network. In general, it is recommended to use larger convolutional filters if the input size is large, skip connections in segmentation tasks, and batch normalization for faster convergence and to obtain better performance \citep{Van2018segmentation}. Ideally, any change made to a standard architecture should be thoroughly explained and reasoned, and the performance improvements should be verified through ablation experiments or comparative studies between custom and conventional architectures.

Pre-trained networks are widely employed but although pre-training is known to improve convergence speed significantly, it might not always lead to a better performance compared to a network trained from scratch, given enough time for convergence \citep{Liu2017, He2018RethinkingIP}. In pre-trained networks, natural scene image databases (e.g., ImageNet) are commonly used, and it is possible that the learned feature representations may not be accurate for histopathology images. Given a training dataset with few images, training only a few of the last decision layers (i.e., freezing the initial layers), using a nonlinear decision layer (i.e., composed of one or more hidden layers with nonlinear activations), using regularization techniques such as weight decay and dropout with ratio $p \in [0.2, 0.5]$ are recommended to avoid overfitting \citep{Tabibu2019, Valkonen2019}.

Until very recently, traditional image processing techniques such as density-based models or handcrafted features were competitive against CNNs in digital histopathology \citep{sharma2017deep}, however, as neural networks have become more and more capable, state-of-the-art results in digital pathology have overwhelmingly come from CNN based methods. It is, however, not entirely clear how much of this increase can be attributed to the most recent advancements in neural networks, as opposed to proper validation, data mining and processing practices, or the general familiarity of researchers with DNN. 

As such comparative studies have yet to exist for digital histopathology, we examined various public histopathology challenges (Refer, Table \ref{Tab:Summary of publicly available datasets}) to assess the impact of the architecture, and found that in many tasks, the specific architecture was not a determining factor in the task objective outcome. For instance, in BACH challenge on breast histology images \citep{Aresta2019}, the winning entry won the 1$^{st}$ place (with a 14\% margin compared to the next best entry) despite using significantly less data without any ensembling networks and a smaller contextual window (i.e., the patch size of the input image), while employing a hard example mining scheme, which made it possible for a network to learn from few examples to converge faster and to avoid overfitting. The winning entry from a MoNuSeg (multi-organ nucleus segmentation) challenge \citep{Kumar2019} employed a UNet without any post-processing step (e.g., watershed transformation or morphological operations to separate nuclei), whereas, the participants using cascaded UNets, ResNets and feature pyramid networks (FPN) or DenseNets consistently scored lower. The winning entry for a TUPAC (tumour proliferation rate estimation challenge) used a hard negative mining technique with a modified ResNet with 6 or 9 residual blocks, and an SVM (support vector machine) for the decision (feature aggregation) layer for mitosis detection \citep{VETA2019111}, beating architectures including GoogleNet, UNet and VGGNet. The winning entry for a CAMELYON16 challenge, including the detection of lymph node metastases \citep{Bejnordi2017} utilized an ensemble of two GoogleNet networks that has given superior results compared to pathologists with time constraints. In contrast, the same network architecture without any stain standardization or data augmentation achieved 10\% and 7\% lower in free-response receiver operator characteristic curve (FROC) and area under curve (AUC) metrics, respectively. Three out of five of the top results used GoogleNet architecture with 7 million parameters (22 layers), and the second best entry employed a ResNet with 44 million parameters (101 layers). Results from a subsequent CAMELYON17 challenge involving detection of cancer metastases in lymph nodes and lymph node status classification \citep{Bandi2018} suggest that using ensemble networks may help self-correct predictions by a suitable form of voting between ensemble networks. In this challenge, the top entry achieved around 2\% better in quadratic weighted Cohen’s kappa metric \citep{cohen1960coefficient} over the second best entry. The top two entries both used ResNet-101 networks, where the top entry used an ensemble of three, and the second best participant used a single network with image resolution four times smaller than the first entry, with about four times the patch size ($960\times960$ versus $256\times256$ pixels).

It is noteworthy that all of the ``\textit{winning networks}" for the various challenges described above were invented on or before the year 2015, whereas challenge dates vary from 2016 to 2019. While challenge scores are not necessarily indicative of the use case performance, as the challenge participants tend to heavily fine-tune their models to achieve the highest possible score, these results indirectly indicate that simpler networks can still prevail, provided that appropriate training practices are applied for the specific problem at hand.

\subsection{Challenges in histopathology image analysis}
\label{ssec:Challenges in histopathology image analysis}

Standard DL architectures require their inputs (e.g., images) in a specific format with certain spatial dimensions. Furthermore, these architectures are generally designed for RGB images, whereas in digital histopathology, working with images in grayscale, HSV or HED color spaces may be desirable for a specific application. Converting images between color spaces, resizing images to fit into GPU memory, quantizing images from a higher bit representation into a lower one, deciding the best resolution for the application at hand and tiling, are some of the choices researchers need to make that will lead to varying degrees of information loss. A reasonable data processing strategy aims to achieve minimal information loss while utilizing architectures to their maximal capacity.

In most applications, it is inevitable that input images will need to be tiled or resized. Memory and computational constraints also make it necessary to find a balance between the required context and the magnification and, as CNNs learn more quickly from smaller images, one should not use images larger than the required context. The optimum trade-off between field of view (FOV) and resolution will depend on the application; for example classifying ductal carcinoma (DCIS) requires a large context to capture morphology, whilst for nuclei detection it is common to use the highest possible power as the required context is as small as one nucleus. In some cases, both high resolution and large FOV are required, for example in cellularity assessment a high power is needed to differentiate between malignant and benign nuclei and a larger FOV is needed to provide the context \citep{akbar2019automated}. A considerable amount of work has been done to combine low and high-resolution inputs in making better decisions in various forms and problems \citep{Li2019,chang2017unsupervised,Wang2019rmdl,li2018based}. However, it is still unclear that these methods are more effective in segmentation tasks compared to selecting a single ``best fit" resolution \citep{seth2018automated}. 

Image pre- and post-processing can be used to boost the task performance of a DL model. In addition to standard preprocessing practices (e.g., resizing an input image and normalization, noise removal, morphological operations to smooth the segmentation masks), preprocessing can also be used to eliminate the need for computationally costly post-processing steps such as iterative refinement of boundaries of segmented regions using conditional random fields \cite{xu2016gland}. \iffalse For instance, \cite{xu2016gland} used an FCN and generated an edge map from HED channel of colon histology images for gland instance segmentation, in addition to the original gland segmentation mask. This is done in order to explicitly learn the second channel of edge or boundary information that can be used to segment glands into multiple instances. This demonstrates that even though each gland mask consists of the boundary information (as its boundaries are part of the mask), an explicit learning task provided by a cost-free relabeling preprocessing step can improve the task performance. Furthermore, relabeling can enable training for different tasks by simply transforming the original data to accommodate the new task. \fi Post-processing techniques can also be used to iteratively refine the model outputs to marginally improve the task objective. Methods based on CRFs are commonly employed to refine boundaries in segmentation tasks (e.g., nuclei) for better delineation of structure boundaries \citep{Qu2019weakly}. Post-processing can also be used for bootstrapping, where the trained model is used for selecting \textit{hard} examples from an unseen test set in which the model underperforms. Then, the model is trained with a subset of original training data and the hard examples obtained from the post-processing step. This form of post-processing is especially useful in selecting a small subset of data from the majority class to prevent class imbalance, or balance the foreground and background samples (i.e., hard negative mining), and is applicable in many tasks including multiple instance learning or segmentation \citep{li2019deep, Kwok2018}. In digital histopathology, hard negative mining is generally used for sampling the normal, or the healthy tissue to avoid over- or under-sampling different types of background regions.

\subsection{Quality of training and validation data}
\label{ssec:Quality of training data}

The success of DL depends on the availability of high-quality training sets to achieve the desired predictive performance \citep{madabhushi2016image,bera2019artificial,niazi2019digital}. 

It is evident from this survey that a vast majority of methods are based on fully-supervised learning. Obtaining a well-curated data set is, however, often expensive and requires significant manual expertise to obtain clean and accurate annotations. There will always be variability between pathologists so ideally the inter-observer agreement should be quantified \citep{Bulten2019automated, akbar2019automated,seth2018automated} and if possible, a consensus between pathologists reached \citep{veta2015assessment}. Some attempts have been made to generate additional annotated data by using alternative techniques like data augmentation \citep{tellez2019quantifying}, image synthesis \citep{hou2019robust} and crowdsourcing \citep{Albarqouni2016}, but it is not yet clear that they are appropriate for digital pathology. In some cases, it is possible to acquire additional information to provide definitive ground truth labels, for example, cytokeratin-stained slides were used to resolve diagnostic uncertainty in the CAMELYON16 challenge \citep{Bejnordi2017}. It is important for researchers to understand how labels are generated and to have some measure of label accuracy.

One way to increase model robustness and improve generalization ability is to include diversity in the training data such as images from multiple scan centres \citep{Campanella2019clinical}, images containing heterogeneous tissue types \citep{hosseini2019atlas} with variations in staining protocols \citep{Bulten2019epithelium}. For instance, \cite{Campanella2019clinical} trained their DL model on an extensive training set containing more than 15,000 patients of various cancer types, obtained across 45 countries. The authors achieved an excellent performance of AUC greater than 0.98 for three histology task, which demonstrates the importance of a large diverse dataset on model performance. With an increase in the number of well-curated open-source datasets hosted by the Cancer Genome Atlas \citep{TCGA}, the Cancer Imaging Archive \citep{TCIA} and various Biomedical Grand Challenges (Refer, Table \ref{Tab:Summary of publicly available datasets}), it is increasingly possible to test methods on a standard benchmark dataset. There is, however, a need for more clinically relevant datasets which capture the complexity of real clinical tasks. The expansion of the breast cancer metastases dataset, CAMELYON16, to CAMELYON17 provides a good illustration of how much larger datasets are needed to assess an algorithm in a more meaningful clinical context \citep{litjens20181399}; in CAMELYON16 399 WSIs from 2 centres were made available but slides containing only isolated tumour cells were excluded and only slide level labels were provided; in CAMELYON17 an additional 1000 WSIs were added from 500 patients and five centres and the total dataset grew to 2.95 terabytes. Even this large dataset does not capture the scale of the clinical task where patients may have multiple WSIs from many more lymph nodes (the CAMELYON set excludes patients with $>5$ dissected nodes), and it also excludes patients who have undergone neoadjuvant therapy which is known to adversely affect classification accuracy \citep{Campanella2019clinical}. 

As the number of clinical centres adopting a fully digital workflow increases, it is likely that the expectation will be that all digital pathology models should be trained and tested on large, clinically relevant datasets. Making such large datasets of WSIs and associated clinical data available publicly poses significant challenges and one way of addressing this may be to move away from the current approach of moving data to the model, and instead, to create mechanisms for researchers (and companies) to move the training and testing of models to the data. A recent example of this was the DREAM mammography challenge \citep{Dream}, where only a small subset of data was released to allow developers to test software, and developers then had to submit docker containers to a central server to access the primary dataset for training and testing. 

\subsection{Model interpretability}
\label{ssec:Model interpretability}

In recent years, DL based methods have achieved near human-level performance in many different histology applications \citep{Campanella2019clinical, Noorbakhsh2019, Coudray2018}, however, the main issue with DL models is that they are generally regarded as a ``black box". Interpretability is less important when networks are carrying out tasks such as mitosis detection or nuclear pleomorphism classification since a pathologist can readily validate performance by visual inspection. Survival models based on a small number of image features that are familiar to pathologists may, therefore, be more acceptable to clinicians than end-to-end deep survival models where it is difficult to understand how a particular prediction is made \citep{holzinger2017we}.  Consequently, several \textit{explainable} AI systems \citep{samek2017explainable, chen2019looks} have been developed, which attempt to gain deeper insights into the working of DL models. In histology, interpretability of DL models has been addressed by using visual attention maps \citep{Huang2019celnet, Bentaieb2018}, saliency maps \citep{tellez2019neural}, heatmaps \citep{paschali2019deep} and image captioning \citep{Zhang2019, weng2019multimodal} techniques. These methods aim to highlight discriminative evidence locations in WSIs by providing pathologists with more clinically interpretable results. For instance, \cite{Zhang2019} presented a biologically inspired multimodal DL model capable of visualizing learned representations to produce rich interpretable predictions using network visual attention maps. Furthermore, their model also learns to generate diagnostic reports based on natural language descriptions of histologic findings in a way understandable to a pathologist. Such multimodal models trained on metadata (such as pathology images, clinical reports and genomic sequences) have the potential to offer reliable diagnosis, strong generalizability and objective second opinions, while simultaneously encouraging consensus in routine clinical histopathology practices. 

One of the most overlooked issue with current DL models is the vulnerability to adversarial attacks \citep{papernot2017practical}. Several recent studies \citep{finlayson2019adversarial,jungo2019assessing,ma2019understanding} have demonstrated that DL system can be compromised by carefully designed adversarial examples, i.e., even small imperceptible perturbations can deceive neural networks in predicting wrong outputs with high certainty. This behaviour has raised concerns in successful real-time integration of these DL systems in critical applications like face recognition \citep{sharif2016accessorize}, autonomous driving \citep{eykholt2017robust} and medical diagnosis \citep{ma2019understanding}. 

Uncertainty maps have been used to identify the failure points of neural networks, and to increase the model interpretability \citep{devries2018leveraging, jungo2019assessing}. In histopathology, uncertainty estimates can also be used to identify rare findings in a slide (e.g., locating lymphoma through high uncertainty regions given a breast cancer metastasis classifier), or as a signal for human interference in labeling the low confidence regions (e.g., active learning), which the network is uncertain \citep{rkaczkowski2019, Graham2019mild}.

\subsection{Clinical translation}
\label{ssec:Clinical translation}

There has been a rapid growth in artificial intelligence (AI) research applied to medical imaging, and its potential impact has been demonstrated by applications which include detection of breast cancer metastasis in lymph nodes \citep{steiner2018impact}, interpreting chest X-rays \citep{nam2018development}, detecting brain tumours in MRI \citep{kamnitsas2016deepmedic}, detecting skin cancers \citep{esteva2017dermatologist}, diagnosing diseases in retinal images \citep{gulshan2016development}, and so on. Despite this impressive array of applications, the real and impactful deployment of AI in clinical practice still has a far way to go. 

The main challenges and potential implications in transforming AI technologies from research to clinical use are as follows. First, the major bottleneck is the regulatory and privacy concern in getting ownership of the patient data such as images and personal health records \citep{bera2019artificial, kelly2019key}. This makes it challenging to train, develop and test safe AI solutions for clinical use. Furthermore, the comparison of DL algorithms in an objective manner is challenging due to variability in design methodologies, which are specifically targeted for a small group of populations. To make fair comparisons, the AI models need to be tested on the same independent test set, which represents the same target population with similar performance metrics. Second, most AI algorithms suffer from inapplicability outside of the training domain, algorithmic bias and can be easily fooled by adversarial attacks \citep{kelly2019key} or by the inclusion of disease subtypes not considered during training. These issues can be partly addressed by developing ``interpretable" AI systems \citep{liu2018artificial, rudin2019stop} which provide a reliable measure of model confidence and also generalization to different multi-cohort datasets. Developing human-centred AI models that can meaningfully represent clinical knowledge and provide a clear explanation for model prediction to facilitate improved interactions with clinicians and machines is of paramount importance. Finally, the algorithms need to be integrated into the clinical workflow. This may be the biggest challenge as very few hospitals have made the significant investment required to implement a  fully digital workflow, which means that microscope slides are not routinely scanned. Transitioning to a digital workflow does, however, result in significant improvements in turnaround time and cost savings \citep{Hanna2019Implementation}, and this is helping to drive increased adoption of digital pathology. If the above challenges are taken into consideration while designing AI solutions, then they are most likely to be transformational in routine patient health care system. 

\section{Conclusions}
\label{sec:Conclusions}

In this survey, we have presented a comprehensive overview of deep neural network models developed in the context of computational histopathology image analysis. The availability of large-scale whole-slide histology image databases and recent advancements in technology have triggered the development of complex deep learning models in computational pathology. From the survey of over 130 papers, we have identified that the automatic analysis of histopathology images has been tackled by different deep learning perspectives (e.g., supervised, weakly-supervised, unsupervised and transfer learning) for a wide variety of histology tasks (e.g., cell or nuclei segmentation, tissue classification, tumour detection, disease prediction and prognosis), and has been applied to multiple cancer types (e.g., breast, kidney, colon, lung). The categorization of methodological approaches presented in this survey acts as a reference guide to current techniques available in the literature for computational histopathology. We have also discussed the critical analysis of deep learning architectures on task performance, along with the importance of training data and model interpretability for successful clinical translation. Finally, we have outlined some open issues and future trends for the progress of this field.

\vspace{0.15cm}
\begin{flushleft}
\textbf{Conflict of interest} 
\end{flushleft} \vspace{-0.5mm}
ALM is co-founder and CSO of Pathcore. CS and OC have no conflicts.

\vspace{0.15cm}
\begin{flushleft}
\textbf{Acknowledgment} 
\end{flushleft} \vspace{-0.5mm}
This research is funded by: Canadian Cancer Society (grant number 705772); National Cancer Institute of the National Institutes of Health [grant number U24CA199374-01]; Canadian Institutes of Health Research.

%%Harvard
\bibliographystyle{model2-names.bst}\biboptions{authoryear}  % add number to see # of references
\bibliography{refs}

\end{document}